\documentclass[a4paper]{article}

\usepackage{amsmath,amsthm,amssymb,mathtools}
\usepackage{comment}
\usepackage[T1]{fontenc} 
\usepackage[mathcal]{euscript}
\usepackage{stmaryrd,graphicx}
\usepackage{cite}

\usepackage[hidelinks]{hyperref}
\usepackage[top=3cm,bottom=3cm,left=2cm,right=2cm]{geometry}
\usepackage{tikz}
\usepackage[export]{adjustbox}
\usepackage{appendix}
\usepackage{setspace}
\usepackage{tcolorbox}
\usepackage{multirow}

\numberwithin{equation}{section}

\usepackage{subcaption}
\usepackage{caption}
\theoremstyle{plain}

\theoremstyle{definition}
\newtheorem{remark}{Remark}[section]

\renewcommand{\d}{\textup{d}}

\makeatletter
\g@addto@macro\bfseries{\boldmath}
\makeatother

\makeatletter
\newcommand{\fixed@sra}{$\vrule height 4\fontdimen22\textfont4 width 0pt\rightarrow$}
\newcommand{\shortarrow}[1]{%
  \mathrel{\text{\rotatebox[origin=c]{\numexpr#1*45}{\fixed@sra}}}}
\makeatother

\usetikzlibrary{shapes.geometric,decorations.markings,intersections,calc,positioning}

\tikzset{->-/.style={decoration={
  markings,
  mark=at position .6 with {\arrow{>}}},postaction={decorate}}}

\tikzset{-<-/.style={decoration={
  markings,
  mark=at position .6 with {\arrow{<}}},postaction={decorate}}}

\tikzset{%
    add/.style args={#1 and #2}{
        to path={%
 ($(\tikztostart)!-#1!(\tikztotarget)$)--($(\tikztotarget)!-#2!(\tikztostart)$)%
  \tikztonodes},add/.default={.2 and .2}}
}  

\tikzset{
    extended line/.style={shorten >=-#1,shorten <=-#1},
    extended line/.default=1cm]
}

\tikzset{line through/.style args={#1 parallel to line through #2 and #3 and
length #4}{insert path={%
let \p1=($(#3)-(#2)$),\n1={atan2(\y1,\x1)} in (#1) -- ++ (\n1:#4)}}}

\definecolor{col1}{rgb}{0.4, 0.69, 0.2}
\definecolor{col2}{rgb}{0.96, 0.29, 0.54}

\definecolor{green(ryb)}{rgb}{0.4, 0.69, 0.2}
\definecolor{frenchrose}{rgb}{0.96, 0.29, 0.54}
\definecolor{persianblue}{rgb}{0.11, 0.22, 0.73}
\definecolor{jade}{rgb}{0.0, 0.66, 0.42}
\definecolor{limegreen}{rgb}{0.2, 0.8, 0.2}

\title{Modified Painlev\'e systems with meromorphic solutions \\ for polynomial Hamiltonians of all degrees} 
\author{Marta Dell'Atti{\color{jade}$\,^1$} \\[-.5ex] \small{\url{m.dell-atti@uw.edu.pl}} \\[1ex] Thomas Kecker{\color{jade}$\,^2$} \\[-.5ex]  \small{\url{thomas.kecker@port.ac.uk}} \\[1ex]
{\color{jade}$\,^1$}{ \small Faculty of Mathematics Informatics and Mechanics, University of Warsaw, Poland} \\[.5ex]
{\color{jade}$\,^2$}{ \small School of Mathematics and Physics, University of Portsmouth, UK}
}
\date{}

\begin{document}
\maketitle

\begin{abstract}
    \noindent We review non-autonomous Hamiltonian systems, polynomial in two dependent variables, with the property that all of their solutions are meromorphic functions in the complex plane. These are related to known Hamiltonian systems with the Painlev\'e property, for which the solutions are single-valued outside a set of fixed singularities. Our systems are equivalent to them in the absence of fixed singularities, and give modified Painlev\'e equations otherwise.
    Using the geometric approach by computing the Okamoto's spaces of initial conditions for certain Hamiltonian systems with general coefficient functions, we obtain differential constraints on these functions for the systems to have only meromorphic solutions. 
    Guided by the Newton polygon of the Hamiltonian function, we obtain all such systems with polynomial Hamiltonian of degree three, four, five, and seven, up to affine equivalence in the dependent variables, while there are none for degree six or degree higher than seven. We thus obtain a list of 12 standard polynomial Hamiltonians that can serve as reference for the Painlev\'e equivalence problem. This list contains also some new Hamiltonians not previously written down, such as quartic Hamiltonians for Painlev\'e I and II, quartic Hamiltonians for the modified Painlev\'e III and V equations, a quintic Hamiltonian for Painlev\'e IV and quintic and septic Hamiltonians for a modified Painlev\'e VI equation.
\end{abstract}

\section{Introduction}

In the quest of defining new transcendental functions as solutions of differential equations, those equations stand out for which all solutions are of meromorphic nature in the complex plane. This is motivated by generalising special functions known from mathematical physics, such as Laguerre, Hermite or Legendre functions, which are given by solutions of certain second-order linear differential equations, as well as elliptic functions. In general, the solutions of nonlinear differential equations, even if analytic locally defined, may develop various types of {\it movable singularities} when continued in the complex plane. These are a non-analytic points of the solution whose locations are determined by the initial conditions for the equations, rather than the equations itself being singular at that point (e.g.\ by some non-autonomous coefficient of the equation being ill-defined at that point). E.g., if the initial conditions prescribed for the equation are varied, the positions of movable singular points for solutions with nearby initial conditions change continuously. While solutions of linear differential equations do not have movable singularities, solutions of nonlinear equations in general have infinitely many of them. This article is concerned with certain nonlinear systems of differential equations that have the property that all their solutions in the complex plane are globally meromorphic functions. In particular, they have the Painlev\'e property, for which all movable singularities of all solutions are poles. While the Painlev\'e equations can exhibit fixed singularities, we consider modified Painlev\'e equations for which fixed singularities are excluded.  We consider polynomial Hamiltonian systems whose solutions are globally meromorphic functions, avoiding fixed singularities by keeping the coefficients of the highest degree terms of the Hamiltonian constant. In this way, we obtain non-autonomous systems with the Painlev\'e property, but the equations reduce to modified Painlev\'e equations in the case where the standard Painlev\'e equation would have fixed singularities. We can relate the two types of equations by a change of variables of the form $t=\exp(z)$, or similar.

The six Painlev\'e equations have a long history since their discovery by P. Painlev\'e and his school at the start of the twentieth century~\cite{Painleve1900,Gam}. In their original form, they were found in the search for new transcendental (meromorphic) solutions for nonlinear differential equations that could not be solved in terms of other known special functions at the time. Painlev\'e classified second-order equations
\begin{equation*}
    \frac{\d^2 w}{\d t^2} = R\!\left(w\,,\frac{\d w}{\d t}\,;t\right),\qquad t \in \mathbb{C} \setminus \Phi, \quad w(t) \in \mathbb{C} \,,
\end{equation*}
where the right-hand side is a rational function in the first two arguments with $t$-dependent coefficients that are also required to be rational. I.e., the coefficients are analytic outside a finite set $\Phi$ of fixed singularities where the solution can be singular, but we require that all solutions are free from movable critical points (such as movable algebraic or logarithmic branch points, as well as essential singularities). The six Painlev\'e equations are listed in the following: 
\begin{align}
\label{Painleve1}
    \text{P}_{\text{I}}\colon~~&\,  \frac{{\rm{d}}^2w}{{\rm{d}}t^2}  =  6\,w^2 + t \,,\\[2ex]
\label{Painleve2}
    \text{P}_{\text{II}}\colon~~&\,  \frac{{\rm{d}}^2w}{{\rm{d}}t^2}  =  2\,w^3 + tw + \alpha \,, \\[2ex]
\label{Painleve3}
    \text{P}_{\text{III}}\colon~~&\,  \frac{{\rm{d}}^2w}{{\rm{d}}t^2}  =  \frac{1}{w} \!\left(\frac{{\rm{d}}w}{{\rm{d}}t}\right)^{\!\!2} \! - \frac{1}{t} \frac{{\rm{d}}w}{{\rm{d}}t} + \frac{1}{t}(\alpha\, w^2 + \beta) + \gamma\, w^3 + \frac{\delta}{w}\,, \\[.7ex]
\label{Painleve4}
    \text{P}_{\text{IV}}\colon~~&\,  \frac{{\rm{d}}^2w}{{\rm{d}}t^2}  =  \frac{1}{2\,w}\!\left(\frac{{\rm{d}}w}{{\rm{d}}t}\right)^{\!\!2} \! + \frac{3}{2} \,w^3 + 4\,t\,w^2 +2(t^2-\alpha)w + \frac{\beta}{w}\,, \\[.7ex]
\label{Painleve5}
    \text{P}_{\text{V}}\colon~~&\,  \frac{{\rm{d}}^2w}{{\rm{d}}t^2}  = \left( \frac{1}{2w}+\frac{1}{w-1} \right) \!\left(\frac{{\rm{d}}w}{{\rm{d}}t}\right)^{\!\!2} \!  - \frac{1}{t} \frac{{\rm{d}}w}{{\rm{d}}t} + \frac{(w-1)^2}{t^2} \left( \alpha w + \frac{\beta}{w} \right) + \frac{\gamma\, w}{t} + \frac{\delta\, w(w+1)}{w-1} \,,\\[.7ex]
    \begin{split} 
\label{Painleve6}
 \hspace*{-5ex}
   \text{P}_{\text{VI}}\colon~~&\,  \frac{{\rm{d}}^2w}{{\rm{d}}t^2}  =   \frac{1}{2}\left( \frac{1}{w} + \frac{1}{w-1} + \frac{1}{w-t} \right) \!\left(\frac{{\rm{d}}w}{{\rm{d}}t}\right)^{\!\!2} \!  - \left( \frac{1}{t} + \frac{1}{t-1} + \frac{1}{w-t} \right) \frac{{\rm{d}}w}{{\rm{d}}t}  \\[.7ex] & \qquad 
    + \frac{w(w-1)(w-t)}{t^2(t-1)^2} \left( \alpha  + \beta\, \frac{t}{w^2} + \gamma\, \frac{t-1}{(w-1)^2} + \delta\, \frac{t(t-1)}{(w-t)^2} \right),
    \end{split} 
\end{align}
with $\alpha,\beta,\gamma,\delta$ constant complex parameters. The properties and certain solutions of these equations have been studied widely by now, see e.g.~\cite{GromakLaineShimomura+2002} for a classic function theoretic treatment or~\cite{fokas2006painleve} for the Riemann-Hilbert approach. While Painlev\'e gave necessary conditions for his equations to have no movable critical points, proofs that the Painlev\'e equations have the Painlev\'e property are by now abundant, starting with a proof for Painlev\'e I by Hukuhara~\cite{Huku}, by Hikkannen and Laine for Painlev\'e I and II~\cite{hikka}, Steinmetz~\cite{Steinmetz} and Hu and Yan~\cite{HuYan} for Painlev\'e I, II and IV, and  Shimomura~\cite{Shimomura2003} for all six Painlev\'e equations. 
There are also the proofs employing the isomonodromy method, see e.g.~\cite{fokas2006painleve}, exploiting the fact that the Painlev\'e equations are the isomonodromic deformation equations for certain linear $2\times 2$ matrix systems.

Note that all six Painlev\'e equations can be written as Hamiltonian systems, studied extensively in a series of papers by Okamoto~\cite{okamoto1,okamoto2,okamoto3,okamoto4,okamoto5}. The so-called Okamoto Hamiltonians are listed below:
\vspace*{-2ex}
\begin{align} 
\label{eq:Okamoto_H1}
H_{\text{I}}^{\text{Ok}}\big(p,q;t\big) &= \frac{1}{2} q^2 - 2 p^3 - tp \,,\\
\label{eq:Okamoto_H2}
H_{\text{II}}^{\text{Ok}}\big(p,q;t\big) &= \frac{1}{2} q^2 - \left( p^2 + \frac{t}{2} \right) q - \kappa \,p \,, \\
\label{eq:Okamoto_H3}
H_{\text{III}}^{\text{Ok}}\big(p,q;t\big) &= \frac{1}{t} \left[ 2 q^2 p^2 - \left( 2 \eta_\infty\, t p^2 + (2\kappa_0 + 1) p - 2 \eta_0\, t \right)q + \eta_\infty (\kappa_0 + \kappa_\infty)tp \right] \,, \\[1ex]
\label{eq:Okamoto_H4}
H_{\text{IV}}^{\text{Ok}}\big(p,q;t\big) &= 2pq^2 - \left(p^2 + 2tp + \kappa_0 \right)q + \kappa_\infty p \,,\\[1ex]
\label{eq:Okamoto_H5}
H_{\text{V}}^{\text{Ok}}\big(p,q;t\big) &= \frac{1}{t} \left[ p(p-1)^2 q^2 - \left( \kappa_0 (p-1)^2 + \kappa_t\, p(p-1) - \eta\, tp \right) q + \kappa (p-1) \right] \,,\\[1ex]
\begin{split} 
\label{eq:Okamoto_H6}
H_{\text{VI}}^{\text{Ok}}\big(p,q;t\big) &= \frac{1}{t(t-1)} \left[ p(p-1)(p-t) q^2 - \left[ \kappa_0\, (p-1)(p-t) + \kappa_1\, p(p-t) \right. \right. \\ & \left. \left. ~~+  (\kappa_t-1)\,p(p-1) \right] q + \kappa(p-t) \right],
\end{split} 
\end{align}
with the parameters $\kappa, \kappa_0, \kappa_1, \kappa_\infty, \kappa_t ,\eta_0, \eta_\infty$ complex parameters related to $\alpha, \beta, \gamma, \delta$ appearing in the standard form of Painlev\'e equations.  

Okamoto also introduced the \emph{space of initial conditions} for each of the six equations, which is an augmented phase space that is uniformly foliated by the solutions~\cite{Okamoto1979}. The existence of such a regular space of initial conditions could also be seen as the defining property for systems with the Painlev\'e property. The systems derived from the (non-autonomous) Hamiltonians $H_{\text{I}}^{\text{Ok}}$ -- $H_{\text{VI}}^{\text{Ok}}$ in the variables $(p(t),q(t))$ lead, under elimination of $q$, to the equations $\text{P}_{\text{I}}$ -- $\text{P}_{\text{VI}}$ in $w=p(t)$. These Hamiltonians, however, are not unique, i.e.\ there usually exist various Hamiltonians leading to the same second-order equation. In this paper, we exploit the existence of a space of initial conditions to recover many of the known Painlev\'e Hamiltonian systems, but also find a few (to our knowledge) new systems -- all of them are bi-rationally related to some known system.

Equivalent Hamiltonians can be identified by the surface type of their spaces of initial conditions, as given in the classification by Sakai~\cite{Sakai2001} of differential and discrete Painlev\'e equations. Using the Okamoto-Sakai theory of Painlev\'e equations, the methods are developed further in~\cite{Kajiwara2017} to provide a geometric framework to explicitly compare equivalent Painlev\'e systems. Recently, e.g.\ in~\cite{Dzhamay2109.06428}, various polynomial and rational Hamiltonians for Painlev\'e IV, V and VI are matched using the geometric approach, which consists of comparing the irreducible components of the inaccessible divisor obtained in the blow-up process when constructing the space of initial conditions. This approach allows one to find bi-rational, symplectic changes of variables relating any pair of Hamiltonian systems belonging to the same type of Painlev\'e equation. 
The geometric approach has also been applied successfully to systems more general than the Painlev\'e equations, such as certain systems of quasi-Painlev\'e type in our earlier papers~\cite{MDAKec,MDAKec2B}, as well systems of Bureau type in~\cite{MDGF1}. 

The geometric approach provides a method to tackle the Painlev\'e equivalence problem in the sense of Clarkson~\cite{Clarkson2019}: Given a system that we suspect to have the Painlev\'e property, how can we decide to which of the Painlev\'e equations it belongs, that is, in terms of which of the Painlev\'e transcendents can one express the solutions of the system? In the present article we answer this question as far as polynomial Hamiltonian systems are concerned, by systematically providing a complete list of polynomial Painlev\'e Hamiltonian systems for all degrees, up to affine equivalence within that degree. While there also exist various rational Hamiltonians for the Painlev\'e equations, such as those in~\cite{ItsPro,JimboMiwa,Kajiwara2017}, even in the polynomial class we find a few, to our knowledge, new cases that have not been considered elsewhere in this form: quartic Hamiltonians for Painlev\'e I and II, quartic Hamiltonians for the modified Painlev\'e III and V equations, a quintic Hamiltonian for Painlev\'e IV, as well as quintic and septic Hamiltonians for a modified Painlev\'e VI equation.

\subsection*{Overview of the article}

The point of the present article is to demonstrate that we can directly derive Hamiltonian systems of Painlev\'e type using the geometric framework. By starting from a general Hamiltonian, with arbitrary analytic coefficient functions, constructing the space of initial conditions leads to a set of differential constraints that need to be satisfied for the solutions of the system to uniformly foliate the space of initial conditions. These differential constraints on the one hand fix the coefficient functions in the Hamiltonian up to linear re-scaling of the independent variable and on the other hand introduce integration constants which become parameters in the Hamiltonian. Our approach in this paper is as follows. For a general polynomial Hamiltonian, we associate its Newton polygon. The number $g$ of lattice points in the interior of the polygon corresponds to the  `generic' genus of the algebraic curve defined by the equation $H(x,y;z_0)=\text{const}$, for fixed $z_0$ (for specific values of $z_0$ the genus can be less than~$g$). Here an example of the Newton polygon for a curve of degree $4$ with $g=3$:
\begin{equation}\label{eq:Ham_NP_ex}
H(x,y)=\sum_{j,k} \alpha_{jk}\, x^j\,y^k = \alpha_{31}\,x^3y+\alpha_{03}\,y^3+\alpha_{20}\,x^2+\alpha_{11}\,xy+\alpha_{02}\,y^2+\alpha_{10}\,x+\alpha_{01}\,y\,, \quad
\includegraphics[width=.14\textwidth,valign=c]{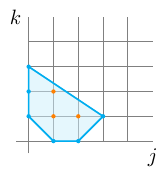} 
\end{equation}
We select those Hamiltonians whose Newton polygons have exactly one interior point, corresponding to the case of elliptic curves. This is related to the fact that solutions of Painlev\'e equations are asymptotic to elliptic functions in certain sectors of the complex plane, whereas the higher genus case leads to equations with movable algebraic singularities, of quasi-Painlev\'e type~\cite{MDAKec2B}. Thus, by selecting those Hamiltonians with one interior point in their Newton polygon we identify the candidates of systems with meromorphic solutions. 
Here we display some examples of the Newton polygons associated with the Hamiltonians of degree $3$ and $4$, by selecting the appropriate diagonal in the lattice:

\vspace*{-3ex}

\begin{equation*}
\arraycolsep=10pt\def\arraystretch{2.2}
\begin{array}{c c c c}
\text{cubic} & \text{cubic} & \text{quartic} & \text{quartic}\\
    \includegraphics[width=.14\textwidth]{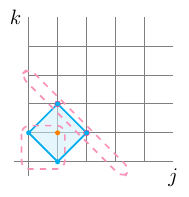} & \includegraphics[width=.14\textwidth]{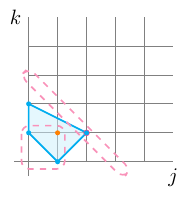} &
    \includegraphics[width=.14\textwidth]{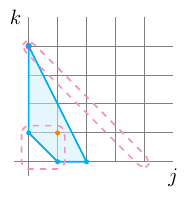} & 
    \includegraphics[width=.14\textwidth]{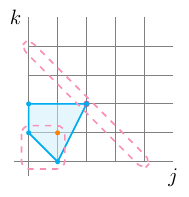}
\end{array} 
\end{equation*}

\vspace*{-1ex}

\noindent
Note that the constant term in the Hamiltonian is not relevant, as it disappears in the derived equations of motion. However, all our Hamiltonians include the linear terms, as these are always present if we allow for a linear shift in the dependent variables\footnote{We specify that the cubic Hamiltonian for the Painlev\'e I equation is slightly different, being degenerate, and one of the linear terms (either in $x$ or $y$) vanishes (see e.g.~\eqref{eq:cubic_coalescence}).}. Namely, for each instance of a polynomial Hamiltonian with arbitrary analytic coefficients, we reduce it to a standard form under affine equivalence, 
\begin{equation}
\label{eq:affine_reduction}
    x \to \widetilde{x} = a(z) x + b(z) y + c(z),\qquad y \to \widetilde{y} = d(z) x + e(z) y + f(z)\,,
\end{equation}
removing or fixing certain coefficients in the process. E.g., in this way we can bring the highest degree homogeneous terms into a suitable form, with constant coefficient $1$.  Under this reduction, in section~\ref{sec:NP} we will select the Newton polygons of polynomial Hamiltonians with exactly one interior lattice point.  Note that the reduction~\eqref{eq:affine_reduction} happens just at the level of the Hamiltonian and is by no means a canonical transformation. Later, we will compare each of the Hamiltonian systems obtained from our standard forms with the systems derived from the Okamoto Hamiltonians under bi-rational changes of variables $\big(x(z),y(z),z\big) \leftrightarrow \big(q(t),p(t),t\big)$ that are indeed canonical.

The Okamoto's space of initial conditions is obtained by finite chains of bi-rational changes of variables~(\emph{cascades} of blow-ups and blow-downs) with the aim of resolving all indeterminacies (\emph{base points}) in the system of equations. As explained in section~\ref{sec:syst_regularisation}, for the system after the final blow-up in each cascade to become regular, a certain cancellation is necessary, giving the aforementioned differential constraints on the coefficient functions. We therefore also refer to these constraints as \emph{regularising conditions}.

While we recover most of the standard Painlev\'e Hamiltonian systems, we also find several Hamiltonians that are (to our knowledge) new in the literature. Some of these were already investigated in our earlier paper~\cite{MDAKec2B}, where we classify quartic Hamiltonian systems with the quasi-Painlev\'e property, an extension of the Painlev\'e property that also allows for movable algebraic poles to occur in the solutions, with the Painlev\'e Hamiltonian systems occurring as special cases with only ordinary poles in this classification. In this paper, we collect polynomial Painlev\'e Hamiltonians for all degrees, where we find cubic, quartic, quintic and septic (but no sextic) degree Hamiltonians, noting that there cannot exist any higher degree ones: the Newton polygon of any Hamiltonian with degree higher than the septic case, and under the constraint above always includes at least two lattice points, which would give rise to a quasi-Painlev\'e system instead. Furthermore, we identify those systems belonging to the same Painlev\'e equation, but with Hamiltonians of different degree, via bi-rational transformations. There are, of course, rational (non-polynomial) Hamiltonians giving rise to the Painlev\'e equations, of which many are known. All of these could also be identified using the geometric approach.

Going by Sakai's seminal classification of differential and discrete Painlev\'e equations~\cite{Sakai2001}, there are really $8$ different types of differential Painlev\'e equations, distinguished by their respective spaces of initial conditions, identified by a \emph{surface type}. These are the equations associated with extended Dynkin diagrams of type $D_j^{(1)}$, for $j \in \{ 4,5,6,7,8 \}$, as well as $E_j^{(1)}$, with $j \in \{6,7,8\}$ (see page~\pageref{DDiag}).
The equations of type $E_6^{(1)}$, $E_7^{(1)}$ and $E_8^{(1)}$ correspond to the Painlev\'e equations $\text{IV}$, $\text{II}$ and $\text{I}$, respectively. These do not exhibit fixed singularities and their solutions can be analytically continued to meromorphic functions in the whole complex plane. Type $D_4^{(1)}$ corresponds to the Painlev\'e $\text{VI}$ equation,  type $D_5^{(1)}$ to the Painlev\'e $\text{V}$ equation, while types $D_6^{(1)}$, $D_7^{(1)}$ and $D_8^{(1)}$ are historically grouped together as the Painlev\'e $\text{III}$ equation, but really represent three distinct forms of this equations, distinguished by their surface type, and should really be seen as three different equations. In particular, referring to the parameters $\alpha, \beta, \gamma, \delta$ in the standard form of Painlev\'e III in~\eqref{Painleve3} following~\cite{Ohyama} we have: 
\begin{equation}
\begin{split} 
    \text{P}_{\text{III-}6}&\colon ~\eqref{Painleve3} \text{ with }\gamma\delta \neq 0\,, \qquad  \text{P}_{\text{III-}7}\colon ~\eqref{Painleve3} \text{ with }\gamma= 0,~ \alpha \delta \neq 0 \text{ or } \delta = 0,~ \beta \gamma \neq 0\,, \\[1ex]
    \text{P}_{\text{III-}8}&\colon ~\eqref{Painleve3} \text{ with }\gamma= 0 ,~ \delta =0, ~\alpha \beta \neq 0\,. 
\end{split} 
\end{equation}
Note that the Painlev\'e III-7 equation cannot be obtained directly from the Okamoto Hamiltonian~\eqref{eq:Okamoto_H3}, but one needs to consider instead (see~\cite{Ohyama}):
\begin{equation}\label{eq:Okamoto_H3_7}
H_{\text{III-7}}^{\text{Ok}}\big(p,q;t\big) = \frac{1}{t} \left[ 2 q^2 p^2 - \left(  (2\kappa_0 + 1) p - 2 \eta_0\, t \right)q + \kappa\,t p \right] \,,
\end{equation}
This is obtained from~\eqref{eq:Okamoto_H3} by considering 
\begin{equation}
    \eta_{\infty} = \varepsilon\,, \qquad \kappa_{\infty} = \frac{\kappa}{\varepsilon}\,, \qquad \varepsilon \to 0\,, \qquad \eta_{\infty}\kappa_{\infty} = \kappa \,.
\end{equation}

By the approach we are taking, we obtain Hamiltonian systems related to the Painlev\'e Hamiltonian systems, but without fixed singularities (in the cases of $\text{P}_{\text{III}}$, $\text{P}_{\text{V}}$ and $\text{P}_{\text{VI}}$). These give rise to the modified Painlev\'e equations, the solutions of which can be meromorphically continued to the entire complex plane (the universal cover of $\mathbb{C}\setminus\{0\})$, but whose coefficients are no longer rational, involving $\exp(z)$ instead. 

For Painlev\'e $\text{III-6}$ and $\text{III-7}$ these equations are (see \cite[Sec.~5.3.2]{MDAKec2B}):
\begin{align}
   \label{mPIIID6}
   \text{P}_{\text{III-6}}^{\text{mod}}\colon ~~ \frac{{\rm{d}}^2y}{{\rm{d}}z^2} &= \frac{1}{y}\!\left(\frac{{\rm{d}}y}{{\rm{d}}z}\right)^{\!\!2} \! - 2b\, y^2 + 4\, y^3 - \exp(z) \left(1+\,a\right) - \frac{\exp(2z)}{y} \,,\\[1ex]
    \label{mPIIID7}
    \text{P}_{\text{III-7}}^{\text{mod}}\colon ~~  \frac{{\rm{d}}^2y}{{\rm{d}}z^2}&=\frac{1}{y}\!\left(\frac{{\rm{d}}y}{{\rm{d}}z}\right)^{\!\!2} \! - 2\exp(z)\,y^2-\exp(z)(1+\,a)-\frac{\exp(2z)}{y} \,,
\end{align}
where $a, b$ are constants. The parameters appearing in our version of modified Painlev\'e~III-6 in~\eqref{mPIIID6} are related to the parameters $\alpha, \beta, \gamma, \delta$ in its standard form~\eqref{Painleve3} through the relations: 
\begin{equation}
    \alpha = - \frac{b}{2}\,, \qquad \beta = \frac{1+a}{4}\,, \qquad \gamma = 1\,, \qquad \delta = -\frac{1}{4}\,,
\end{equation}
via the mapping (following~\cite{GromakLaineShimomura+2002})
\begin{equation}\label{eq:changeP36_std_mod}
   t = \exp(z/2)\,, \qquad w(t) = y(z)\,\exp(-z/2)\,.
\end{equation}

To obtain the standard version of Painlev\'e III-7 (i.e.~\eqref{Painleve3} with $\gamma=0$) the change of variables from~\eqref{mPIIID7} is simpler, involving the independent variable only, 
\begin{equation}
    t=-\exp(z)\,, \qquad w(t)=y(z)\,. 
\end{equation}
The parameters in~\eqref{mPIIID7} are related to those appearing in the standard form via the mapping:
\begin{equation}
    \alpha = 2\,, \qquad \beta = 1+a\,, \qquad \delta = -1\,. 
\end{equation}

Note that while Painlev\'e III-6 and Painlev\'e III-7 (associated with $D^{(1)}_6$ and $D^{(1)}_7$ respectively) can be obtained from a polynomial Hamiltonian system, Painlev\'e III-8 is special in this regard, as it cannot be obtained from a polynomial Hamiltonian, but rather from a Hamiltonian that is rational in its arguments~\cite{okamoto5}:
\begin{equation}\label{eq:Ham_P38_mod}
H^{\text{Ok}}_{\text{III'-8}}\big( p,q;t \big) = \frac{1}{t} \left( p^2 q^2 + pq \right) -\frac{1}{2}\frac{p}{t} - \frac{1}{2p} \,,
\end{equation}
and the second order equation in $p(t)$ is denoted Painlev\'e III'-8
\begin{equation}
    \text{P}_{\text{III'-8}}: ~~\frac{\d^2 p}{\d t^2} = \frac{1}{p}\left(\frac{\d p}{\d t}\right)^{\!\!2} -\frac{1}{t} \,\frac{\d p}{\d t} + \frac{p^2}{t^2} -\frac{1}{t}\,, 
\end{equation}
that can be mapped into the standard form of Painlev\'e III-8 \eqref{Painleve3} with $\alpha=1$, $\beta=-1$, $\gamma=0$, $\delta=0$ via the change of variables 
\begin{equation}
    t = z^2 \,, \qquad p(t) = z\,y(z)\,,
\end{equation}
in the variable $y(z)$.  Using the methods of the present paper, we obtain a (rational) Hamiltonian (see the remark~\ref{rmk:P38}) leading to a modified equation of type Painlev\'e III-8,
\begin{equation}\label{eq:modifiedP38}
\text{P}^{\text{mod}}_{\text{III-8}}:~~   \frac{\d^2 y}{\d z^2} = \frac{1}{y} \!\left( \frac{\d y}{\d z}\right)^{\!\!2} - 2\exp(z)\, y^2 + 2\,. 
\end{equation}

The modified version of Painlev\'e V is~\cite[p. 29]{GromakLaineShimomura+2002}
\begin{equation}
       \label{mPV}
    \text{P}_{\text{V}}^{\text{mod}}\colon ~~
\frac{\d^2 w}{\d z^2} = \left( \frac{1}{2w} + \frac{1}{w-1} \right) \!\left(\frac{\d w}{\d z}\right)^{\!\!2} \! +(w-1)^2 \left( \alpha\,w+\frac{\beta}{w} \right) + \gamma \, \exp(z)\, w+ \frac{\delta \, \exp(2z) \,w (w+1)}{w-1} \,,
\end{equation}
{and this is related to the standard version of Painlev\'e V~\eqref{Painleve5} via the mapping for the independent variables $\exp(z)=t$. }

Further below, we will also obtain a modified version of the Painlev\'e VI equation from a system with quintic, or respectively septic, Hamiltonian:
\begin{equation}
\label{mP6}
\begin{split}
    &\hspace*{-2ex}{\text{P}}^{\text{mod}}_{\text{VI}}\colon \frac{\d^2 w}{\d z^2}  =  \frac{1}{2}\left(\frac{1}{w}+\frac{1}{w+1}+\frac{1}{w-\exp(z)} \right) \left( \frac{\d w}{\d z}\right)^{\!\!2}+\!\left(\frac{1}{1+\exp(z)}+\frac{w}{\exp(z)-w} \right)\left( \frac{\d w}{\d z}\right) \\[1ex]
    &\,+\frac{w(1+w)(\exp(z)-w)}{2(1+\exp(z))^2} \left(  (1 + \exp(z)) \left(\frac{(b-a+1)^2}{(1 + w)^2} + \frac{
 \left((a_{12}+b)^2 -1 \right)\exp(z)}{(w-\exp(z))^2}  \right) - \frac{a^2\exp(z) }{
 w^2}  -4\, \widetilde{a}_{01}^{\,2}\right) 
\end{split}
\end{equation}
where $a,b,a_{12},\widetilde{a}_{01} \in \mathbb{C}$ are parameters, obtained as integration constants of differential conditions on the coefficient functions in the Hamiltonian system. Our parameters are related to the parameters in equation~\eqref{Painleve6} by
\begin{equation}
\label{mP6coeff}
    \alpha = -4\,\widetilde{a}_{01}^{\,2}, \quad \beta = -a^2, \quad \gamma = (b-a+1)^2, \quad \delta = (a_{12}+b)^2 -1\,.
\end{equation}
Note that, different to the other modified Painlev\'e equations, equation~\eqref{mP6}, does have (infinitely many) fixed singularities, namely where $z \in i \pi + 2 i \pi \mathbb{Z}$.

\section{Painlev\'e Hamiltonian systems and the geometric approach}\label{sec:geom_approach}
Sakai's classification~\cite{Sakai2001}, though applied to difference Painlev\'e equations, has its origin in the construction of the space of initial conditions, performed by K.\ Okamoto for each of the six (differential) Painlev\'e equations~\cite{Okamoto1979}, or rather the Hamiltonian systems giving rise to these equations. The space of initial conditions is an augmented phase space for which the solutions of the equation form a uniform foliation. While Okamoto started from the known Hamiltonian systems with the Painlev\'e property derived by~\eqref{eq:Okamoto_H1}~--~\eqref{eq:Okamoto_H6} and constructed their spaces of initial conditions, one can also obtain the Painlev\'e systems by starting from general Hamiltonians with a priori arbitrary (analytic) coefficient functions and determine the conditions under which the system has a regular space of initial conditions. That is, given a polynomial Hamiltonian function of a certain degree,
\begin{equation}
\label{polyHam}
H^{(\text{deg})}\big(x(z),y(z);z\big) = \sum_{j,k} a_{jk}(z)\, x^j y^k, \qquad \text{deg}=\max (j+k)
\end{equation}
assumed to have analytic coefficient functions $a_{jk}(z)$, one can ask, what are the conditions on these functions if the Hamiltonian system is to have the Painlev\'e property, if at all possible.

To obtain a complete list of Hamiltonian systems with the desired property, we proceed as in one of our previous papers~\cite{MDAKec2B}, where we presented a classification of quasi-Painlev\'e Hamiltonian systems with quartic Hamiltonian, under affine equivalence. The quasi-Painlev\'e property is an generalisation of the Painlev\'e property, where we allow for movable algebraic poles as well as ordinary poles to occur as movable singularities in the solutions. While in the present paper we only consider ordinary poles (i.e.\ the Painlev\'e case), we employ the same two 
mechanisms as in~\cite{MDAKec2B}. Starting from the most general case of a Hamiltonian, with all coefficient functions involved, we produce subcases via \emph{degeneration} and  \emph{coalescence}. Here, degeneration refers to the reduction of genus in the Hamiltonian by setting certain combinations of coefficients that appear in the cascades of blow-ups to $0$, thereby changing the leading-order behaviour of the movable singularities. This is not needed here, as all Hamiltonians we start with already have genus $1$. Coalescence of base points refers to the limiting process where the locations of two previously separate base points are merged, also by setting certain coefficient functions or combinations thereof to zero, or introducing an auxiliary parameter which is sent to $0$ to this effect.
For example, one can coalesce two base points arising in the cascades of blow-ups of the Painlev\'e V equation to first obtain Painlev\'e III-6, then coalesce two further base points in a second cascade to obtain Painlev\'e III-7 (see the figure below and further explanation in section~\ref{sec:quartic}):
\begin{equation}\label{eq:P5_P36_P37}
\begin{array}{c c c c c}
    \includegraphics[width=.14\textwidth,valign=c]{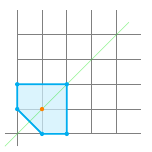} & \xrightarrow[a_{20}(z)\to 0]{\text{coal.}} & 
    \includegraphics[width=.14\textwidth,valign=c]{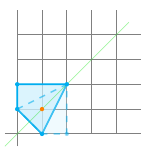} & \xrightarrow[a_{02}(z) \to 0]{\text{coal.}}  & \includegraphics[width=.14\textwidth,valign=c]{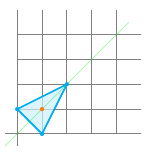} \\[1.5ex]
    \text{\small Painlev\'e V} & &  
    \text{\small Painlev\'e III-6} & &  
    \text{\small Painlev\'e III-7}  
\end{array}
\end{equation}
In this case, the coalescences have the effect of setting the coefficients, first $a_{20}(z)$ and second $a_{02}(z)$ in~\eqref{polyHam} to $0$, changing the shape of the Newton polygon, but not its genus. 
As special cases in our classification of quartic Hamiltonians~\cite{MDAKec2B}, we recovered the Painlev\'e equations $\text{P}_{\text{I}}$, $\text{P}_{\text{II}}$, $\text{P}_{\text{III-6}}$, $\text{P}_{\text{III-7}}$ and $\text{P}_\text{V}$. 
In the way this classification is performed, defining the coefficients of the leading-order (quartic) terms in the Hamiltonian to be constants, the systems we consider naturally have no fixed singularities, i.e.\ their solutions are meromorphic in the entire plane. The systems we find are therefore related to the usual Painlev\'e equations $\text{P}_\text{I}$ and $\text{P}_\text{II}$ on the one hand, and to the modified Painlev\'e equations $\text{P}_\text{III}$ and $\text{P}_\text{V}$ (see e.g.~\cite{GromakLaineShimomura+2002}) on the other. These modified Painlev\'e equations are obtained by the change in dependent variables, $t=\exp(z)$, thus transferring the equations to the universal cover $\exp: \mathbb{C} \to \mathbb{C}\setminus\{0\}$ of the punctured complex plane $\mathbb{C}\setminus\{0\}$, which is the complex plane itself. In the remainder of this paper, we will obtain modified Painlev\'e systems VI, V and IV, with Hamiltonians of degree $\deg=5$ and $\deg=7$, noting that there cannot exist any for other degrees, $\deg=6$ and $\deg>7$ due to the Newton polygons in these cases containing at least two interior lattice points.

\subsection{Surface type}\label{sec:surface_type}
The first step in computing the space of initial conditions is the compactification of the phase space, given a Hamiltonian system derived from a Hamiltonian of the form~\eqref{polyHam} in the variables $(x(z),y(z))\in \mathbb{C}^2$. To analyse the behaviour of the solutions of the system at the movable singularities, where $\max\{|x(z)|,|y(z)|\} \to \infty$, the phase space is extended to a complex projective surface, like $\mathbb{CP}^2$ or $\mathbb{CP}^1 \times \mathbb{CP}^1$. Throughout this work, we will use $\mathbb{CP}^2$ as the initial compactification, represented as  
\begin{equation}\label{eq:CP2}
\includegraphics[width=.2\textwidth,valign=c]{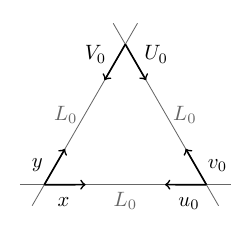} \qquad  
    \begin{aligned}
    & \mathbb{CP}^2 \simeq \mathbb{A}^2_{(x,y)} \cup \mathbb{A}^2_{(u_0,v_0)} \cup \mathbb{A}^2_{(U_0,V_0)}\,, \\[1ex] 
    &[\,1:x:y\,] = [\,u_0:1:v_0\,] = [\,V_0:U_0:1\,]\,, \\[.7ex]
    &u_0 = \frac{1}{x}\,, \qquad V_0 = \frac{1}{y}\,, \qquad v_0 = \frac{y}{x} = \frac{1}{U_0} \,,
\end{aligned}
\end{equation}
which is obtained by glueing three affine charts, here with coordinates $(x,y)$, $(u_0,v_0)$ and $(U_0,V_0)$ respectively, all functions of the independent variable $z \in \mathbb{C}$. The system is then written in the three affine charts $\mathbb{A}^2_{(x,y)}$, $\mathbb{A}^2_{(u_0,v_0)}$ and $\mathbb{A}^2_{(U_0,V_0)}$ 
\begin{equation}\label{eq:systemsP2}
    \begin{cases}
        \dfrac{\d x}{\d z} = f_1(x,y;z) = -\dfrac{\partial H}{\partial y} \\[2ex] 
        \dfrac{\d y}{\d z} = g_1(x,y;z) = \dfrac{\partial H}{\partial x} \\[2ex] 
    \end{cases}\,, \qquad 
    \begin{cases}
        \dfrac{\d u_0}{\d z} = f_2(u_0,v_0;z) \\[2ex] 
        \dfrac{\d u_0}{\d z} = g_2(u_0,v_0;z) \\[2ex] 
    \end{cases}\,, \qquad 
    \begin{cases}
        \dfrac{\d U_0}{\d z} = f_3(U_0,V_0;z) \\[2ex] 
        \dfrac{\d V_0}{\d z} = g_3(U_0,V_0;z) \\[2ex] 
    \end{cases}\,, 
\end{equation}
where the functions $f_1$, $g_1$ are polynomials, while $f_j, g_j$ for $j=2,3$ are in general rational and not directly derived from the transformed Hamiltonian function due to the non-symplectic nature of the transformation -- one would need to multiply the left-hand side by Jacobian of the transformation as e.g.\ explained in~\cite{MDAKec}. 

In each of the three charts, we look for \emph{base points}, where the system's flow is ill-defined due to an indeterminacy in the vector field of the form $0/0$, causing the solutions going through such a point all being tangent (to some higher degree) with one another. To resolve such an indeterminacy, a blow-up transformation is applied, replacing each base point with a curve $\simeq \mathbb{CP}^1$ that represents all the possible directions through which to approach the base point, called the \emph{exceptional curve}. By applying a certain number of blow-ups, we can separate out the different solution curves.
In a coordinate chart $(u_i,v_i)$, a blow-up at a point $p\colon (u_i,v_i)=(a,b)$ is performed by
\begin{equation}
  \text{Bl}_p(\mathbb{C}^2) = \{ (u_i,v_i) \times [w_0:w_1] \in \mathbb{C}^2 \times \mathbb{CP}^1: (u_i-a) w_0 = (v_i-b) w_1 \}\,,
\end{equation}
giving rise to two new coordinate charts, namely
$(u_j,v_j)$, $(U_j,V_j)$, according to the bi-rational transformations
\begin{equation}\label{eq:blowup_coord}
    \begin{cases}
        u_j = u_i - a\\[1ex] 
        v_j = \dfrac{v_i - b}{u_i - a}
    \end{cases}\,, \qquad 
    \begin{cases}
        U_j = \dfrac{u_i - a}{v_i - b}\\[1ex] 
        V_j = v_i - b
    \end{cases}\,.
\end{equation}
The exceptional curve corresponds, as a point set, to a union of sets in the charts $(u_j,v_j)$, $(U_j,V_j)$ 
\begin{equation}
    E = \{ u_j = 0 \} \cup \{ V_j = 0 \},
\end{equation}
also known as the exceptional divisor. A divisor on a smooth surface is a formal sum of irreducible curves (co-dimension $1$ subspaces), and we consider divisors modulo linear equivalence under addition of principle divisors (those appearing as the zero and pole set of meromorphic functions). This equivalence gives rise to the divisor class group, also known as the Picard group under a certain identification. For example, in $\mathbb{CP}^2$ the only generating element of the Picard group of $\mathbb{CP}^2$ is the hyperplane divisor denoted by $L_0$ in~\eqref{eq:CP2} and has self-intersection number $1$, as given by the intersection form $\bullet$. An exceptional divisor (e.g.\ the class of an exceptional curve obtained by a blow-up), is by definition a divisor with self-intersection $-1$. The intersection between $L_0$ and any exceptional curve, as well as the intersection between two different exceptional curves, is $0$. The intersection form on an $n$ times blown up $\mathbb{CP}^2$ is the linear extension of these relations on hyperplane divisor and exceptional divisors $E_1,\dots,E_n$,
\begin{equation}\label{eq:intersection_form}
L_0 \bullet L_0 = 1, \qquad E_i \bullet L_0 = 0 \quad\forall i, \qquad E_i \bullet E_j = -\delta_{ij}\,. 
\end{equation}
The appearance of the exceptional curve $E$ after a blow-up affects the self-intersection of the lines involved according to~\eqref{eq:intersection_form}. 
After a blow-up, new base points may arise on the exceptional curve and usually, to completely remove the indeterminacy, a sequence of finitely many successive changes of coordinates is required. 

Here we have an example of a typical sequence of blow-ups transformations at base points with coordinates $(U_0,V_0)=(0,0)$ with the emergence of the exceptional curve $E_1$, at coordinate $(U_1,V_1)=(0,0)$ producing $E_2$, and at $(u_2,v_2)=(0,0)$ with the emergence of $E_3$: 
\vspace*{-2ex}
\begin{equation*}
\begin{array}{c c c c}
\hspace*{-2ex}
    \includegraphics[width=.16\textwidth,valign=c]{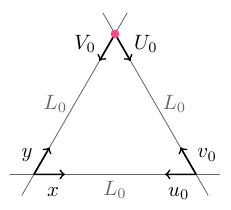}  \leftarrow \hspace*{-2ex}
    &\includegraphics[width=.22\textwidth,valign=c]{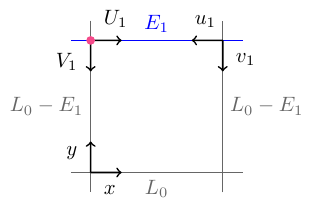}  \leftarrow \hspace*{-2ex}
    &\includegraphics[width=.25\textwidth,valign=c]{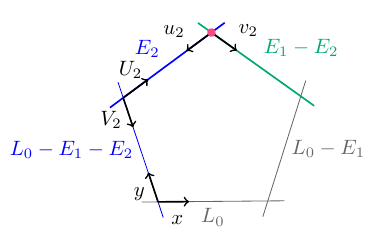} \leftarrow \hspace*{-2ex}
    &\includegraphics[width=.28\textwidth,valign=c]{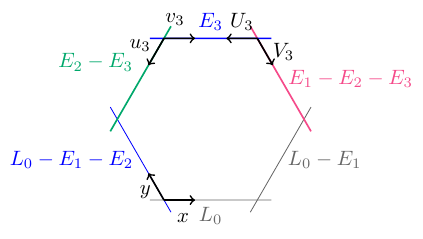} \\ 
    (a) & (b) & (c) & (d)
\end{array} 
\end{equation*}
In each digram, the red dot locates the coordinates of the base point; the curves depicted in blue exhibit self-intersection $-1$, those in green $-2$ and in red $-3$; the curves in gray have self-intersection $\ge 0$. 
As mentioned above, the occurrence of exceptional curves modifies the self-intersection value of the adjacent curves, according to the form~\eqref{eq:intersection_form}. E.g., in $(c)$ after the emerging of $E_2$, the curve labelled by $E_1-E_2$ has self-intersection $(E_1-E_2)\bullet (E_1-E_2)=-2$. At the end of the finite sequences (cascades) of transformations originating from the base points in $\mathbb{CP}^2$, we produce a (not necessary minimal) rational surface, depicted by the configuration of the components of the exceptional divisor. 

Whenever in presence of a redundant component (not necessary in the regularisation process), we apply the reverse operation of the blow-up (represented by $\leftarrow$): the blow-down (represented by $\to$). While a blow-up resolves a singularity by replacing a point with an exceptional curve, a blow-down contracts such a curve back to a point. This operation is aimed to obtain the minimal configuration of curves in the rational surface. For each of the six Painlev\'e equations the minimal configuration is uniquely determined by a specific configuration of the $-2$-curves identified in the rational surface. The dual graph of the specific configuration gives a particular extended Dynkin diagram of type $D^{(1)}_j$ or $E^{(1)}_j$, in this framework called \emph{surface type}: 
{\small
    \begin{equation*}\label{DDiag}
\arraycolsep=15pt\def\arraystretch{2.2}
    \begin{array}{c c c c c}
    \includegraphics[width=.045\textwidth]{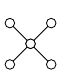} &\includegraphics[width=.07\textwidth]{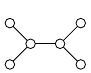} &\includegraphics[width=.09\textwidth]{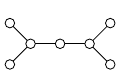} &\includegraphics[width=.11\textwidth]{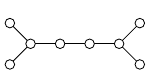} 
    &\includegraphics[width=.13\textwidth]{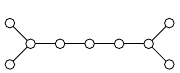} 
    \\[-2ex]
    D_4^{(1)},\hyperref[Painleve6]{\text{ P}_{\text{VI}}} & D_5^{(1)},\hyperref[Painleve5]{\text{ P}_{\text{V}}} & D_6^{(1)},\hyperref[Painleve3]{\text{ P}_{\text{III-6}}} & D_7^{(1)},\hyperref[Painleve3]{\text{ P}_{\text{III-7}}} & D_8^{(1)},\hyperref[Painleve3]{\text{ P}_{\text{III-8}}}
    \\[1ex] 
     \includegraphics[width=.09\textwidth]{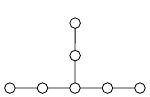} 
     &\includegraphics[width=.14\textwidth]{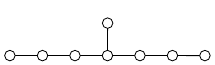}
     &\includegraphics[width=.17\textwidth]{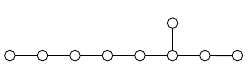}
     & & \\[-2ex]
    E_6^{(1)},\hyperref[Painleve4]{\text{ P}_{\text{IV}}} & E_7^{(1)},\hyperref[Painleve2]{\text{ P}_{\text{II}}} & E_8^{(1)},\hyperref[Painleve1]{\text{ P}_{\text{I}}} & &
    \end{array}
    \end{equation*}}

The Okamoto's space of initial conditions is then defined as the maximal space on which the flow of the vector field (i.e.\ the solutions of the equation) yield a uniform foliation. That is, it is the extended, $n$-times blown up and $m$ times blown down phase space from which we remove the components of the exceptional divisor that are inaccessible by the flow of the vector field (also called the inaccessible divisor). In the case of the Painlev\'e systems compactified on $\mathbb{CP}^2$, only the difference $n-m = 9$ is fixed.

\subsection{System regularisation}\label{sec:syst_regularisation}
Starting from the systems~\eqref{eq:systemsP2} in the affine charts of $\mathbb{CP}^2$, derived from the Hamiltonian~\eqref{polyHam} with generic coefficient functions $a_{jk}(z)$, we remove each base point occurring in these systems through a cascade of blow-ups. This is achieved through successive application of transformations of the form~\eqref{eq:blowup_coord}, with $i=j+1$. We here demonstrate this procedure for the simple case of a Hamiltonian of cubic degree\footnote{The label of the Hamiltonian is the same of that introduced in section~\ref{sec:cubic}.}: 
\begin{equation}
\begin{aligned} 
H_2^{(3)}(x(z),y(z);z) &= xy^2 + x^2 + b_{10}(z) \,x + b_{01}(z) \,y \,. 
\end{aligned}
\end{equation}
The system derived from this Hamiltonian is regularised by two cascades with $6$ and $3$ blow-ups, respectively, starting at the origins of the affine charts $\mathbb{A}_{(u_0,v_0)}$ and $\mathbb{A}_{(U_0,V_0)}$ in~\eqref{eq:CP2}. The long cascade is:
\begin{equation}\label{eq:cubic_casc_1}
\begin{split} 
     (u_0,v_0) = (0,0) \,\,&\leftarrow\,\,  (U_1,V_1)=(0,0)\,\, \leftarrow\,\,  \left(\widetilde{u}_2\equiv \frac{1}{u_2}, \widetilde{v}_2 \equiv v_2 \right) = (0, -1)  \,\,\leftarrow\,\, (u_3, v_3) = (0,0)\,\,\leftarrow \\ &\leftarrow\,\, 
     (u_4,v_4)=(0,\,-b_{10}(z))\,\, \leftarrow\,\, (u_5, v_5) = \left(0,\,-b_{01}(z)+{\dfrac{\d b_{10}}{\d z}}\right),
    \end{split}
\end{equation}
whereas the short cascade is given by
\begin{align}\label{eq:cubic_casc_2}
    (U_0,V_0) =(0,0) \,\,&\leftarrow\,\,  (U_7,V_7)=(0,0)\,\, \leftarrow\,\,  (U_8, V_8) = (-b_{01}(z), 0)  \,.
\end{align}
After the final blow-up in each cascade the resulting system of differential equations is free from base points. However, for generic functions $b_{01}(z)$, $b_{10}(z)$ the systems do not give a regular initial value problem on the exceptional curve from the final blow-up in a cascade. E.g.\ the system in the final chart of the cascade in~\eqref{eq:cubic_casc_1} takes on the form
\begin{equation}
\label{uv6systems}
\frac{\d u_6}{\d z} = P_6(u_6,v_6;z) , \qquad \frac{\d v_6}{\d z} = \frac{1}{u_6}\left(\frac{\d^2 b_{10}}{\d z^2}-\frac{\d b_{01}}{\d z} + u_6\, Q_6(u_6,v_6;z)\right),
\end{equation}
where $P_6$, $Q_6$ are polynomials in the first two arguments.  
This system only becomes regular under the condition 
\begin{equation}
\frac{\d^2 b_{10}}{\d z^2} -\frac{\d b_{01}}{\d z}=0\,,
\end{equation}
when an additional cancellation of $u_6$ occurs. Similarly, the cascade~\eqref{eq:cubic_casc_2} leads to the system 
\begin{equation}
\label{UV9system}
    \frac{\d U_9}{\d z} = \frac{1}{V_9}\left( \frac{\d b_{01}}{\d z} + V_9\, P_9(U_9,V_9;z) \right), \quad \frac{\d V_9}{\d z} = Q_9(U_9,V_9;z)\,,
\end{equation}
where $P_9$, $Q_9$ are polynomials in the first two arguments. 
The system becomes regular under the condition 
\begin{equation}
    \frac{\d b_{01}}{\d z}=0\,. 
\end{equation}
Thus, each cascade of blow-up gives rise to a differential constraint among the coefficient functions, and together they give:  
\begin{equation}
    \frac{\d b_{01}}{\d z} =0 \,, \qquad \frac{\d^2 b_{10}}{\d z^2}=0\,. 
\end{equation}
With these conditions implemented, $b_{01}$ is a constant while $b_{10}(z) = az+b$ is at most linear in $z$. In case $a\neq0$, shifting and re-scaling $z$ so that $b_{10}(z)=z$, the second-order equation satisfied by $y$ is the standard Painlev\'e II equation~\eqref{Painleve2}. If $a=0$, the Hamiltonian system becomes autonomous and can be integrated by quadrature, resulting in elliptic functions as solutions.

With the additional cancellation, we see that the systems~\eqref{uv6systems} and~\eqref{UV9system}, by Cauchy's local existence and uniqueness theorem, have analytic solutions at any point $z_* \in \mathbb{C}$ for initial conditions $(u_6(z_*),v_6(z_*)) = (0,h)$ on the exceptional curve $u_6=0$, respectively $(U_9(z_*),V_9(z_*))=(k,0)$ on the exceptional curve $V_9=0$. Here $h$, respectively $k$, refer to the position on the exceptional curve where the solution intersects the curve. These give series solutions, e.g.\
\begin{equation}
    u_6(z) = (z-z_*) + \mathcal{O}\big((z-z_*)^2\big), \qquad v_6(z) = h + z_*(1-b_{01}) (z-z_*) + \mathcal{O}\big((z-z_*)^2\big)\,,
\end{equation}
which, under all the changes of variables made in the blow-up process, translate back into a simple pole at $z_*$ of the solution in the original variables $(x(z),y(z))$.

In the remainder of this paper, for every Hamiltonian considered we obtain, for each cascade of blow-ups, one differential constraint. When solved simultaneously, the resulting equations give rise to systems with meromorphic solutions which are related to either Painlev\'e or modified Painlev\'e equations.

\subsection{Newton polygons}\label{sec:NP}

The classification in~\cite{MDAKec2B} lists the possible types of quasi-Painlev\'e systems for polynomial Hamiltonians up to degree $4$ in $x$ and $y$ and presents some standard form for each type, reduced under affine equivalence, together with the differential constraints to be satisfied by the coefficient functions in the Hamiltonian.
While the type of the equation is determined by the surface diagram of the equation, we also associate the Newton polygon of the Hamiltonian to each type of equation, as shown in equation~\eqref{eq:Ham_NP_ex}. Guided by the principle of selecting only Newton polygons with exactly one interior point, we identify all different forms of Painlev\'e systems with polynomial Hamiltonian. While most of them are known previously, there are a few (to our knowledge) new cases such as a quartic Painlev\'e I and II Hamiltonian, as well as a quintic Painlev\'e IV Hamiltonian, and we present them here together: for each equation (surface type) and each degree of the Hamiltonian we will give a standard form, up to equivalence under affine transformations in $x$, $y$. For Hamiltonians with different degree, but for the same equation (surface type), there exist bi-rational transformations between the corresponding equations of motion. Below we give a list of all possible degrees of Hamiltonians, and the representation of the Newton polygon associated with them, with the constraint of always containing the linear terms in both the variables $x$ and $y$: 
{\small
\begin{equation*}\arraycolsep=3pt
\begin{array}{c c c c c c c c}
 \multicolumn{2}{c}{\text{Painlev\'e I}} & \quad  & \multicolumn{2}{c}{\text{Painlev\'e II}} & \quad  & \text{Painlev\'e III-6} & \text{Painlev\'e III-7} \\[2ex] 
 \includegraphics[width=.14\textwidth]{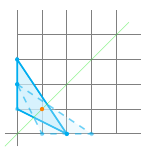} & \includegraphics[width=.14\textwidth]{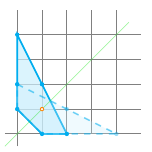} & & 
 \includegraphics[width=.14\textwidth]{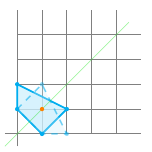} & \includegraphics[width=.14\textwidth]{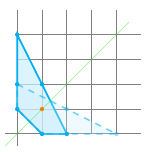} & &
 \includegraphics[width=.14\textwidth]{NP_P3_6_quartic.pdf} & \includegraphics[width=.14\textwidth]{NP_P3_7_quartic.pdf} \\
 \text{cubic }\hyperref[C3_ham]{(H^{(3)}_3)} & \text{quartic }\hyperref[eq:Ham_P1_quartic]{(H^{(4)}_5)} & & \text{cubic }\hyperref[C2_ham]{(H^{(3)}_2)} & \text{quartic }\hyperref[eq:Ham_P2_quartic]{(H^{(4)}_4)} & & \text{quartic }\hyperref[quartic_Ham_4_3]{(H^{(4)}_3)} & \text{quartic }\hyperref[quartic_Ham_4_2]{(H^{(4)}_2)} \\[6ex]
 \multicolumn{2}{c}{\text{Painlev\'e IV}} & & \multicolumn{2}{c}{\text{Painlev\'e V}} & & \multicolumn{2}{c}{\text{Painlev\'e VI}} \\[2ex] 
 \includegraphics[width=.14\textwidth]{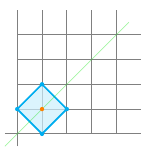} & \includegraphics[width=.14\textwidth]{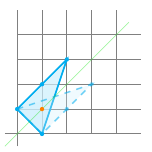} & &
 \includegraphics[width=.14\textwidth]{NP_P5_quartic.pdf} & \includegraphics[width=.14\textwidth]{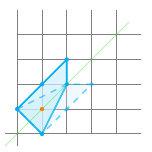} & &
 \includegraphics[width=.14\textwidth]{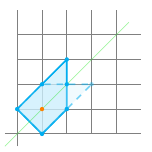} & \includegraphics[width=.14\textwidth]{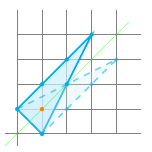}\\
 \text{cubic }\hyperref[C1_ham]{(H^{(3)}_1)} & \text{quintic }\hyperref[H5_P4_simp]{(H^{(5)}_3)} & & \text{quartic }\hyperref[quartic_Ham_4_1]{(H^{(4)}_1)} & \text{quintic }\hyperref[H5_P5]{(H^{(5)}_2)} & & \text{quintic }\hyperref[quintic_Ham_P6_modified]{(H^{(5)}_1)} & \text{septic }\hyperref[eq:H7_explicit]{(H^{(7)}_1)} \label{NP_H7} 
\end{array}    
\end{equation*}
}

One can convince oneself, more or less by inspection, that these are the only possible ways of enclosing exactly one point in the interior of a Newton polygon, given the corners $(1,0)$ and $(0,1)$. 
The cubic Painlev\'e~I, II and IV polygons correspond to the standard Okamoto Hamiltonians~\eqref{eq:Okamoto_H1},~\eqref{eq:Okamoto_H2} and~\eqref{eq:Okamoto_H4}. The quartic Painlev\'e~III-(6/7) polygons are correspond to different choices of parameters in the Okamoto Hamiltonian~\eqref{eq:Okamoto_H3} (the Painlev\'e III-8 equation cannot be obtained in this way). As already mentioned above, the quartic Painlev\'e~I, II, quartic modified Painlev\'e III and V, and quintic Painlev\'e IV cases are new in this respect. The quartic Painlev\'e V Hamiltonian, already obtained in~\cite{MDAKec2B}, is also related to the Hamiltonian given by Kajiwara-Noumi-Yamada in~\cite{Kajiwara2017}, while the quintic Painlev\'e V case is related to the standard Okamoto Hamiltonian~\eqref{eq:Okamoto_H5} for Painlev\'e V. The quintic Painlev\'e VI case is related to the Okamoto Hamiltonian~\eqref{eq:Okamoto_H6}. 
We note that, while there is no possible polygon for a sextic Hamiltonian with one interior point, there is exactly one polygon relating to a septic Hamiltonian. This also corresponds to the  surface type $D_4^{(1)}$ (Painlev\'e VI), and is mentioned already in~\cite{KimuraMatuda1980}, where it is also attributed to Okamoto.

\section{Summary of cubic and quartic Painlev\'e Hamiltonians}
For each degree $d \geq 3$ of the polynomial Hamiltonian, using the geometric approach we will identify all cases leading to systems with the Painlev\'e property, up to linear equivalence in the dependent variables. In the cubic case, we recover the Hamiltonians belonging to the Painlev\'e equations $\text{P}_{\text{I}}$, $\text{P}_{\text{II}}$ and $\text{P}_{\text{IV}}$, while in the quartic case we obtain Hamiltonian systems for the modified Painlev\'e equation $\text{P}_{\text{V}}^{\text{mod}}$ (type $D_5^{(1)}$), and modified $\text{P}_{\text{III}}$ (types $D_6^{(1)}$ and $D_7^{(1)}$), as well as Hamiltonians for $\text{P}_{\text{I}}$ and $\text{P}_{\text{II}}$ (already presented in~\cite{MDAKec2B}).

To avoid fixed singularities in the equations, we take the highest degree terms in the Hamiltonian to have constant coefficients, while all other coefficients in principle are analytic functions, but these will be restricted by conditions for the equations to have the Painlev\'e property. The conditions are obtained in constructing the space of initial conditions.

\subsection{Cubic case}\label{sec:cubic}
In the cubic case, we find the usual Hamiltonians associated with the Painlev\'e equations I, II and IV. This is seen as follows. We fix the coefficients of the homogeneous terms of degree $3$ in the Hamiltonian to be constant and factor it as
\begin{equation}
\begin{aligned}\label{eq:H3_hom}
H^{(3)}_{\text{hom}}(x,y;z) &= a_{30}\, x^3 + a_{21}\, x^2y + a_{12}\, xy^2 + a_{03}\, y^3 = \prod_{i=1}^3 \left( \alpha_i \,x + \beta_i \,y \right), \quad \alpha_i, \beta_i \in \mathbb{C}\,, 
\end{aligned} 
\end{equation}
we identify linear factors under the equivalence relation
\begin{equation}
    (\alpha_i,\,\beta_i) \sim (\lambda \,\alpha_i, \lambda \,\beta_i)\,, \quad  \lambda \in \mathbb{C}\,, 
\end{equation}
that is we consider the coefficients of these factors as points $[\alpha_i:\beta_i] \in \mathbb{CP}^2$, $i=1,2,3$, corresponding to three base points 
\begin{equation}\label{eq:H3_base_points}
p_1: (u_0,v_0)=(0,0), \qquad p_2: (u_0,v_0) = (0,1), \qquad p_3: (U_0,V_0)=(0,0)\,, 
\end{equation}
in the extended phase space of this Hamiltonian system. By a M\"obius transformation, these points can be fixed to be $[1:0]$, $[1:1]$ and $[0:1]$, corresponding to the homogeneous part $x(x-y)y$ in~\eqref{eq:H3_hom}. 

\subsubsection{Painlev\'e IV}

Through an additional shift in $x$ and $y$ we can let the coefficients $a_{20}(z)=a_{02}(z)= 0$, obtaining the general Hamiltonian:
\begin{equation}
\label{C1_ham}
H^{\text(3)}_1\big(x(z),y(z);z\big) = x(x-y)y + a_{11}(z) \,xy + a_{10}(z) \,x + a_{01}(z) \,y \,.
\end{equation}
Performing the blow-up process on the three base points, we obtain the regularisation conditions
\begin{equation}
    \frac{\d^2 a_{11}}{\d z^2} = 0\,, \qquad \frac{\d a_{10}}{\d z} = 0\,, \qquad \frac{\d a_{01}}{\d z} = 0\,,
\end{equation}
i.e.\ $a_{11}(z) = az + b$ and $a_{10},a_{01} \in \mathbb{C}$ constants. In the case $a\neq0$, we can shift and re-scale $z$ so that $a_{11}(z) = z$. {In this way, the system associated with this Hamiltonian is related to the system derived from the} Okamoto Hamiltonian $H^{\text{Ok}}_{\text{IV}}(p,q;t)$ in~\eqref{eq:Okamoto_H4} via the following expressions: 
\begin{equation} 
    z=\sqrt{2}\,t\,,\qquad x(z) = \sqrt{2}\, q(t)\,, \qquad y(z) = \frac{p(t)}{\sqrt{2}}\,,  
\end{equation}
and the relations for the parameters {
$\kappa_0=-2\,a_{10}$, $\kappa_\infty=-a_{01}$.}
The Painlev\'e IV~\eqref{Painleve4} equation is satisfied by the variable $p(t)$. 
\begin{remark}
A slightly different normalisation for the homogeneous cubic part of the Hamiltonian is suggested in~\cite{Kecker2019}, leading to 
$$H = \frac{1}{3} \left(x^3 + y^3 \right) + z \, xy + a_{10} x + a_{01} y\,,$$ 
for which the Painlev\'e IV equation is satisfied for $w = x+y-z$.    
\end{remark}

\subsubsection{Painlev\'e II}

Under the coalescence of base points $p_2 \to p_3$ in~\eqref{eq:H3_base_points}, the Hamiltonian is now reduced to the form:
\begin{equation}
\label{C2_ham}
H^{\text(3)}_2\big(x(z),y(z);z\big) = xy^2 + x^2 + a_{10}(z)\, x + a_{01}(z) \,y \,,    
\end{equation}
and one obtains the regularisation conditions 
\begin{equation}
    \frac{\d^2 a_{10}}{\d z^2}=0\,,\qquad\frac{\d a_{01}}{\d z}=0\,,
\end{equation}
i.e.\ $a_{10}=az+b$ and $a_{01} \in \mathbb{C}$. Again, when $a\neq 0$, we shift and re-scale $z$ so that $a_{10}(z) = z$. The Hamiltonian { system derived from}~\eqref{C2_ham} then can be directly linked to the Hamiltonian {system derived from} $H^{\text{Ok}}_{\text{II}}(p,q;t)$ in~\eqref{eq:Okamoto_H2} with the following transformations: 
\begin{equation}
    z= -2^{-1/3} t\,, \qquad x(z)=-2^{-1/3}\,q(t)\,, \qquad y(z) = 2^{1/3}\,p(t)\,, 
\end{equation}
{and the parameters relation $\kappa=-a_{01}$,
yielding Painlev\'e II for $p(t)$.}

\subsubsection{Painlev\'e I}

Under a further coalescence of base points $p_1 \to p_3$ in~\eqref{eq:H3_base_points}, we obtain the Hamiltonian:
\begin{equation}
\label{C3_ham}
H^{\text(3)}_3\big(x(z),y(z);z\big) = y^3 + x^2  + a_{01}(z) \,y\,,
\end{equation}
for which the only regularisation condition is
\begin{equation}
    \frac{\d^2 a_{01}}{\d z^2}=0\,, 
\end{equation}
i.e.\ $a_{01}=az + b$. Again in this case, if $a\neq 0$ the variable $z$ can be shifted and re-scaled, so that $a_{01}(z)=z$, and the { Hamiltonian system associated with the} standard Okamoto Hamiltonian $H_{\text{I}}^{\text{Ok}}(p,q;t)$ in~\eqref{eq:Okamoto_H1} is recovered implementing the transformations
\begin{equation}
    z = 2^{-1/5}\,t\,, \qquad x(z)=-2^{-2/5}\,q(t)\,,  \qquad y(z) = -2^{2/5}\,p(t)\,. 
\end{equation}
Thus, one can derive the second order equation for $p(t)$, i.e.~Painlev\'e I.

The two subsequent coalescences of the base points $p_1, p_2, p_3$ in~\eqref{eq:H3_base_points} are here represented in terms of the respective Newton polygons:
\begin{equation}\label{eq:cubic_coalescence}
\begin{array}{c c c c c}
\includegraphics[width=.14\textwidth,valign=c]{NP_P4_cubic.pdf} & \xrightarrow[p_2 \to p_3]{\text{coal.}} & 
\includegraphics[width=.14\textwidth,valign=c]{NP_P2_cubic.pdf} & \xrightarrow[p_1 \to p_3]{\text{coal.}}  & \includegraphics[width=.14\textwidth,valign=c]{NP_P1_cubic.pdf} \\[1.5ex]
    \text{\small Painlev\'e IV} & &  
    \text{\small Painlev\'e II} & &  
    \text{\small Painlev\'e I}  
\end{array}
\end{equation}

For each of the systems derived from the Hamiltonians~\eqref{C1_ham}, \eqref{C2_ham}, \eqref{C3_ham}, if $a\neq0$, one can shift and re-scale $z$ so that $a_{11}, a_{10}$ and $a_{01}$ are $z$ respectively. 
In case $a=0$ one obtains autonomous systems that can be solved in terms of elliptic functions. In this paper, we are mainly interested in the non-autonomous case.

\subsection{Quartic case}
\label{sec:quartic}
In the case of a quartic Hamiltonian we will see in the following that we encounter Painlev\'e Hamiltonian systems belonging to the (modified) Painlev\'e equations $\text{I}$, $\text{II}$, $\text{III-6}$, $\text{III-7}$ and $\text{V}$. These are obtained as special cases in the more general classification of quasi-Painlev\'e systems~\cite{MDAKec2B}. 
Analogously to the cubic case, the homogeneous terms in the Hamiltonian of degree $4$ are fixed to a constant and factorised as follows: 
\begin{equation}
    H^{(4)}_{\text{hom}}(x,y;z) = a_{40}\,x^4 + a_{31}\,x^3y+ a_{22}\,x^2y^2 + a_{13}\,xy^3+ a_{04}\,y^4 = \prod_{i=1}^4 (\alpha_i\, x+ \beta_i \,y)\,, \qquad \alpha_i, \beta_i \in \mathbb{C}\,,
\end{equation}
with the coefficients $[\alpha_i\colon \beta_i] \in \mathbb{CP}^2$, $i=1, \dots, 4$ locating the four base points, 
\begin{equation}\label{eq:H4_base_points}
p_1: (u_0,v_0)=(0,0), \quad p_2: (u_0,v_0) = (0,1), \quad p_3: (U_0,V_0)=(0,0), \quad p_4: (U_0,V_0)=(k,0),
\end{equation}
that can be associated with the positions $[1:0]$, $[1:1]$, $[0:1]$ and $[0:k]$, with $k \in \mathbb{C}\setminus \{0,1\}$ a parameter. In this way, the homogeneous term of degree $4$ becomes $xy(x-y)(x-ky)$, and the general form for the Hamiltonian of degree $4$ is: 
\begin{equation}
\begin{split} 
 \text{Q1}:\quad  H^{(4)} &= xy(x-y)(x-ky) + a_{21}(z)\,x^2y+a_{12}(z)\,xy^2 + a_{20}(z)\,x^2 + a_{11}(z)\,xy + a_{02}(z)\,y^2 \\
    &~~+ a_{10}(z)\,x + a_{01}(z)\,y \,,   
\end{split}
\end{equation}
labelled with Q1 in~\cite{MDAKec2B}, whose Newton polygon is shown as the first element of the chain in~\eqref{eq:Q1_Q2_Q3_P5}. Subsequent coalescences of base points and degeneracies yield the first Newton polygon of interest here, with genus $1$. In particular, with the labelling of the different cases as in~\cite[Sec.~3.2]{MDAKec2B}: 
\begin{equation}\label{eq:Q1_Q2_Q3_P5}
\arraycolsep=1pt\def\arraystretch{2.2}
\begin{array}{c c c c c c c c c}
\includegraphics[width=.14\textwidth,valign=c]{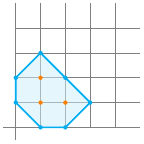} & \xrightarrow[p_4 \to p_3]{\text{coal.}} & 
\includegraphics[width=.14\textwidth,valign=c]{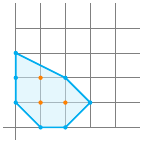} & \xrightarrow[p_2 \to p_1]{\text{coal.}}  & \includegraphics[width=.14\textwidth,valign=c]{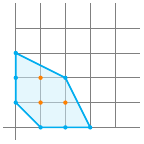} & \xrightarrow[a_{30}\to0]{\text{deg.}}  & \includegraphics[width=.14\textwidth,valign=c]{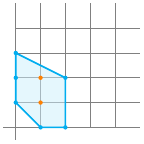} & \xrightarrow[a_{03}\to0]{\text{deg.}} & \includegraphics[width=.14\textwidth,valign=c]{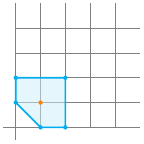} \\[-2ex]
     \text{\small Q1} &  & \text{\small Q2} & & \text{\small Q3}  & & \text{\small Q3.a1} & & \text{\small Painlev\'e V}
\end{array}
\end{equation}
The coalescences of base points in~\eqref{eq:H4_base_points} $p_4 \to p_3$ and $p_2 \to p_1$ lead to a Hamiltonian with leading homogeneous term $x^2y^2$. In this case, through a shift in $x$ and $y$ we can eliminate the terms $x^2y$ and $xy^2$ at the expense of re-introducing terms $a_{30}(z)x^3$ and $a_{03}(z)y^3$, arriving at the Hamiltonian for Q3,
\begin{equation}
\begin{split} 
    \text{Q3}:\quad   H^{(4)} &= x^2y^2 + a_{30}(z)\,x^3+a_{03}(z)\,y^3 + a_{20}(z)\,x^2 + a_{11}(z)\,xy + a_{02}(z)\,y^2
    + a_{10}(z)\,x + a_{01}(z)\,y \,. 
\end{split}
\end{equation}

\subsubsection{Painlev\'e V}

Through two degenerations in the Hamiltonian, by letting first $a_{30}=0$ and second $a_{03}=0$ we arrive at a Hamiltonian with Newton polygon of genus $1$, depicted at the end of~\eqref{eq:Q1_Q2_Q3_P5}:
\begin{equation}
    H^{(4)}_1 \big(x(z),y(z);z \big) = x^2y^2 + a(z)^2(x^2 + y^2) +a_{11}(z)\,xy + a_{10}(z)\, x + a_{01}(z)\,y \,.
\end{equation}
The regularisation process allows us to determine the differential constraints on the coefficient functions
\begin{equation}\label{eq:H4_conds}
    \frac{\d a_{11}}{\d z}=0 \,, \qquad a\,\frac{\d^2 a}{\d z^2}  - \left(\frac{\d a}{\d z}\right)^{\!\!2} = 0\,, \qquad  \frac{\d}{\d z} \left( \frac{a_{10}}{a} \right) = 0\,, \qquad \frac{\d}{\d z} \left( \frac{a_{01}}{a} \right) = 0\,, 
\end{equation}
necessary for cancellations in the differential systems at the end of the four branches belonging to the cascades of the two base points: 
\begin{equation}\label{eq:H4_cascades}
\setlength{\arraycolsep}{0pt}
    \begin{array}{c c c c c c c}
         & & p_2^+~~ \leftarrow ~~ p_3^+ & & & &  p_4^+~~ \leftarrow ~~ p_5^+    \\
         & \shortarrow{5} & & & &\shortarrow{5} & \\[-1ex]
         p_1:(u_0,v_0)=(0,0)& & & \hspace{10ex}   & p_3:(U_0,V_0)=(0,0) & & \\
         & \shortarrow{3} & & & & \shortarrow{3} & \\[-2ex]
         & & p_2^-~~ \leftarrow ~~ p_3^- &  & & &  p_4^-~~ \leftarrow ~~ p_5^-    
    \end{array}
\end{equation}
Solving~\eqref{eq:H4_conds}, we obtain a very symmetric version of a Hamiltonian system giving rise to the modified Painlev\'e~V equation~\eqref{mPV}, symmetric in $x$ and $y$:
\begin{equation}
\label{quartic_Ham_4_1}
    H^{\text(4)}_1\big(x(z),y(z);z\big) = x^2 y^2 - \exp(2z) \left( x^2 + y^2 \right) + c\, xy + \exp(z) (a \,x + b\, y), \quad a,b,c \in \mathbb{C}.
\end{equation}
Note that the equation contains three parameters, $a,b,c \in \mathbb{C}$, while there are four parameters in the second-order Painlev\'e $\text{V}$ equation~\eqref{Painleve5}, one of which is redundant. 
In particular, letting  
\begin{equation}
x(z) = \text{exp}\!\left( \frac{t}{2} \right)\!\!\left(\frac{w(t)+1}{w(t)-1}\right)\,, \qquad z = \frac{t}{2} \,,
\end{equation}
in~\eqref{quartic_Ham_4_1} and eliminating $y(z)$ from the Hamiltonian system obtained, we can deduce the second-order equation for $w(t)$, with the following relations with  parameters $\alpha,\beta,\gamma,\delta$ in the standard Painlev\'e V~\eqref{Painleve5}:
\begin{equation} 
\begin{split}
    \alpha &= \frac{1}{32} \left(1+b+c\right)^2 \,,\qquad  \beta = -\frac{1}{32} \left(1-b+c\right)^2 \,, \qquad  \gamma = a \,,   \qquad  \delta = -2 \,. 
\end{split} 
\end{equation}
To relate the Hamiltonian  system derived from \eqref{quartic_Ham_4_1} with the { one associated with the} Okamoto Hamiltonian $H^{\text{Ok}}_{\text{V}}$ in~\eqref{eq:Okamoto_H5}, we implement the following changes of variables 
\begin{equation}
    \exp(z) = \sqrt{t}\,, ~~ x(z) = \sqrt{t}\,\frac{p(t)+1}{p(t)-1} \,, ~~ y(z) = \frac{1}{\sqrt{t}} \left(\big(p(t)-1\big)^2q(t) + 
  \frac{1 - (a +c)}{4} \big(p(t)-1\big) + t  \right)\,,
\end{equation}
and the relations for the parameters 
\begin{equation}
    \eta=
   2\,, \qquad \kappa=-\frac{(b - a+2) (c+ a-1)}{16} \,, \qquad \kappa_0=
   \frac{c- b+1}{4} \,, \qquad \kappa_t=\frac{a-2}{2} \,.  
\end{equation}
The variable $p(t)$ then satisfies Painlev\'e II. 

\subsubsection{Painlev\'e III}
Under the coalescence of base points $p_2^+ \to p_2^-$ in~\eqref{eq:H4_cascades}, and the implementation of the differential constraints associated with the respective regularisation process, one obtains the Hamiltonian:
\begin{equation}
\label{quartic_Ham_4_2}
    H^{\text(4)}_2\big(x(z),y(z);z\big) = x^2 y^2 - y^2 + a \,xy + \exp(z)\, x+ b\, y \,,
\end{equation}
with two parameters $a,b \in \mathbb{C}$, which can be related to the modified Painlev\'e~III-6 equation~\eqref{mPIIID6} for the variable~$y(z)$. { We can relate the Hamiltonian {system derived from}~\eqref{quartic_Ham_4_2} to {that associated with the} Okamoto Hamiltonian $H^{\text{Ok}}_{\text{III}}$ in~\eqref{eq:Okamoto_H3} for III-6 via the mapping 
\begin{equation}
    \exp(z/2) = t\,, \qquad x(z) = 1-\frac{q(t)}{t} \,, \qquad y(z)= t\,p(t)\,,
\end{equation}
and the relations for the parameters 
\begin{equation}
    \eta_0 = -1\,, \qquad \eta_{\infty} = 1 \,, \qquad \kappa_0 = a\,, \qquad \kappa_\infty = a+2b  \,. 
\end{equation}
}

Under the further coalescence $p_4^+ \to p_4^-$ in~\eqref{eq:H4_cascades} and the respective regularisation process, one produces a Hamiltonian system derived from:
\begin{equation}
\label{quartic_Ham_4_3}
H^{\text(4)}_3\big(x(z),y(z);z\big) = x^2y^2 + a \,xy + \exp(z)\left( x + y \right) \,,
\end{equation}
with only one parameter, $a \in \mathbb{C}$, remaining. Here, both the variables $x(z)$ and $y(z)$ satisfy the modified Painlev\'e~III-7 equation~\eqref{mPIIID7}. { The Hamiltonian system derived from~\eqref{quartic_Ham_4_3} is then mapped into the system derived from the Okamoto Hamiltonian III-7~\eqref{eq:Okamoto_H3_7} via the transformations
\begin{equation}
    \exp(z) = t\,, \qquad x(z) = 2\,q(t)\,, \qquad y(z) = p(t)\,,
\end{equation}
and the relations for the parameters 
\begin{equation}
    \eta_0 = -\frac{1}{2}\,, \qquad \kappa_0 = - \frac{a+1}{2}\,, \qquad \kappa = -\frac{1}{2}\,. 
\end{equation}

} 

\begin{remark}\label{rmk:P38}
    As mentioned in the introduction, the Painlev\'e III-8 equation cannot be obtained from a polynomial Hamiltonian. However, if we apply our method to the rational Hamiltonian
    \begin{equation}
        H^{\text{mod}}_{\text{III-8}}\big(x(z),y(z);z \big)  = x^2 y^2 + a_{11}(z)\, x y + a_{01}(z)\, y + \frac{1}{y}\,, 
    \end{equation}
we find the regularisation conditions
\begin{equation}
    \frac{\d a_{11}}{\d z} = 0, \qquad  a_{01}\,\frac{\d^2 a_{01}}{\d z^2} - \left( \frac{\d a_{01}}{\d z} \right)^{\!\!2} = 0\,,
\end{equation}
implying that $a_{11} \in \mathbb{C}$ is a constant and $a_{01}(z) = \exp(z)$, possibly after re-scaling the independent variable. With these parameters, we recover the modified Painlev\'e III-8 equation~\eqref{eq:modifiedP38}.

\end{remark}

Apart from these systems corresponding to the modified Painlev\'e III and V equations, we also find quartic Hamiltonians for Painlev\'e I and II, as follows. Starting from the Hamiltonian of the case Q3, a different sequence of degenerations and coalescences than in~\eqref{eq:Q1_Q2_Q3_P5} yields the cases relating to Painlev\'e II and I: 
\begin{equation}\label{eq:Q3_Q5_P2_P1}
\arraycolsep=1pt\def\arraystretch{2.2}
\begin{array}{c c c c c c c c c}
\includegraphics[width=.14\textwidth,valign=c]{NP_Q3.pdf} & \xrightarrow[p_3 \to p_1]{\text{coal.}} & 
\includegraphics[width=.14\textwidth,valign=c]{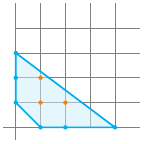} & \xrightarrow[]{\text{deg.}}  & \includegraphics[width=.14\textwidth,valign=c]{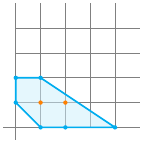} & {\color{red}\xrightarrow[]{\quad}}  & \includegraphics[width=.14\textwidth,valign=c]{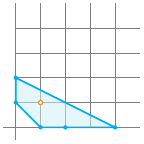} & \xrightarrow[]{\text{coal.}} & \includegraphics[width=.14\textwidth,valign=c]{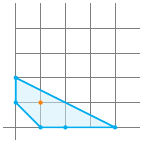} \\[-2ex]
     \text{\small Q3} &  & \text{\small Q5} & & \text{\small Q5.a1}  & & \text{\small Painlev\'e II} & & \text{\small Painlev\'e I}
\end{array}
\end{equation}
Here, under the coalescence of the base points for Q3, i.e.\ $p_3 \to p_1$ in~\eqref{eq:H4_base_points}, the highest degree term in the Hamiltonian becomes $x^4$. 

\subsubsection{Painlev\'e II}
Under two further degenerations of the Hamiltonian we arrive at a Hamiltonian with surface type $E_7^{(1)}$, which in the most general form is
\begin{equation}
H^{(4)}\big( x(z),y(z);z \big)=x^4 +a_{21}(z)\, x^2 y +a_{20}(z)\, x^2 +a_{11}(z)\, x y +a_{02}(z)\, y^2+a_{10}(z)\, x + a_{01}(z)\, y  \,.
\end{equation}
There is, however, some redundancy in the coefficient functions. Through a bi-rational transformation\footnote{This is the only place in this paper where we consider a non-affine, bi-rational transformation to compare Hamiltonians of the same degree. This is what the red arrow in~\eqref{eq:Q3_Q5_P2_P1} represents, a combination of a degeneration and coalescence of base points (see~\cite{MDAKec2B} for the details).} of the form
\begin{equation}\label{eq:transf_std_P2}
x(z)\to a(z)\, x(z)+b(z), \qquad y(z)\to d(z) \,y(z)+e(z)\, x(z)+f(z)\, x(z)^2+g(z)\,,
\end{equation}
with certain functions $a(z),\dots,g(z)$ we can reduce the Hamiltonian to the form (e.g.~\cite[eq.~(3.3)]{MDAKec})
\begin{equation}
    H^{(4)}\big( x(z),y(z);z \big) = x^4 + y^2 + a_{20}(z)\, x^2 + a_{10}(z)\, x\,,
\end{equation}
which, under the conditions $a_{10}(z)=a_{10} \in \mathbb{C}$ and $a_{20}(z) = z$ is a standard Hamiltonian, from which the second-order Painlev\'e II equation~\eqref{Painleve2} is readily derived. A different choice for $a(z),\dots,g(z)$ in the transformation~\eqref{eq:transf_std_P2} leads to the Hamiltonian that we already obtained in~\cite[Sec.~5.5.2]{MDAKec2B}:
\begin{equation}
\label{eq:Ham_P2_quartic}
    H^{\text(4)}_4\big(x(z),y(z);z\big) = x^4  + a_{21}\, x^2 y  + y^2 + a_{10}(z)\, x + a_{01}(z)\, y\,, 
\end{equation}
where $a_{21} \neq \pm 2$.  The regularisation conditions are 
\begin{equation}
    \frac{\d a_{10}}{\d z} = 0, \quad \frac{\d^2 a_{01}}{\d z^2} = 0\,,
\end{equation}
meaning that $a_{10}$ is constant and $a_{01}(z)$ at most linear in $z$, and we let $a_{01}(z)=z$. We get the exact Painlev\'e~II~\eqref{Painleve2} under a linear change in dependent and independent variable,
\begin{equation}
    x(z) = \frac{2a_{21}}{\sqrt{a_{21}^2 - 4}}\, w(t)\,, \qquad z = 2 a_{21}\, t\,, \qquad \alpha = -\frac{(1 + 2a_{10}) \sqrt{a_{21}^2 - 4} }{8a_{21}^3 }\,. 
\end{equation}
{
To relate the Hamiltonian {system derived from}~\eqref{eq:Ham_P2_quartic} with {the system derived from the} Okamoto Hamiltonian $H^{\text{Ok}}_{\text{II}}$~\eqref{eq:Okamoto_H2}, we implement the following change of variables 
\begin{equation}
    z=\frac{k}{k^2-1}\,t\,, \qquad x(z) = p(t)\,, \qquad y(z) = k\,p(t)^2 -\frac{k^2-1}{2k}\,q(t) -\frac{k^3}{(k^2-1)^2}\,t\,,
\end{equation}
and the relation for the parameters 
\begin{equation}
    \kappa = a_{10}-\frac{2K^2}{(K-1)^3}\,,
\end{equation}
where $K \equiv k^2$ is a zero of the cubic 
\begin{equation}
    K^3 -K^2+5K -1 = 0\,. 
\end{equation}
}

\subsubsection{Painlev\'e I}

Due to the condition $a_{21} \neq \pm 2$, the regularisation of the Hamiltonian system~\eqref{eq:Ham_P2_quartic} has two branches of blow-ups. The coalescence of the two branches is effected by letting $a_{21} \to 2$ (or $-2$), and a term $a_{11} x y$ is re-introduced (for details see our earlier paper~\cite{MDAKec2B}). The Hamiltonian of quartic degree is then 
\begin{equation}
\label{eq:Ham_P1_quartic}
    H^{(4)}_5 \big( x(z),y(z);z \big) = x^4 + 2\,x^2y + y^2 +a_{11}\,xy + a_{10}(z)\,x + a_{01}\,y\,, 
\end{equation}
and we determine $a_{10}(z)=z$ by solving the only regularising condition obtained for the Hamiltonian system. To explicitly recover Painlev\'e I~\eqref{Painleve1}, we need to fix all the remaining parameters in the second order equation determined for $x(z)$, and consider linear transformations of the dependent and independent variables: 
\begin{equation}
    x(z) = w(t) \pm \frac{1}{\sqrt{3}}-1 \,, \qquad z = 2t\,, 
\end{equation}
with $a_{11}=4$, $a_{01}=4(2\pm\sqrt{3})$. 
{ To directly relate {the Hamiltonian system derived from}~\eqref{eq:Ham_P1_quartic} with { that associated with the} Okamoto Hamiltonian $H^{\text{Ok}}_{\text{I}}$ in~\eqref{eq:Okamoto_H5}, we implement the changes of variables 
\begin{equation}
\begin{aligned} 
    z = -\frac{1}{2^{1/3}}\,t + \frac{a_{01}}{2^{1/3}}\,,\quad  x(z) &= p(t)\,, \quad y(z) = 2^{2/3}q(t) - p(t)^2-\frac{a_{11}\,p(t)+a_{01}}{2}\,,  
\end{aligned}
\end{equation}
and the relations for the parameters 
\begin{equation}
    a_{01} = - \frac{1}{2^{2/3}} \,, \qquad a_{11} = 2^{2/3}\,.
\end{equation}
The variable $w(t)=p(t)$ thus satisfies the Painlev\'e I equation~\eqref{Painleve1}. 
}

While there is no quartic Hamiltonian for Painlev\'e IV, there is a quintic one that we will derive in the next section, besides Hamiltonians for the modified Painlev\'e $\text{V}$ equation and a modified version of Painlev\'e $\text{VI}$.

\section{Quintic Hamiltonian systems}
\label{quintic}

Starting from a general quintic Hamiltonian,
\begin{equation}
    H^{(5)}\big(x(z),y(z);z\big) = \sum_{0 \leq i+j \leq 5} a_{ij}(z) \,x^i y^j\,,
\end{equation}
one can, in principle, perform a similar analysis as we did in~\cite{MDAKec2B} and determine, through a process of subsequent degenerations and coalescences of base points, all systems of quasi-Painlev\'e type within this class. Here, we are mainly interested in systems with the Painlev\'e property. Rather than performing the complete classification of quintic Hamiltonian systems of quasi-Painlev\'e type and picking out the systems with the Painlev\'e property, which would be a long process, we can directly identify those cases that lead to systems with the Painlev\'e property, by selecting the respective systems corresponding to Newton polygons with exactly one interior point. For this to be the case, the only possibility is for the homogeneous quintic part of the Hamiltonian to be $x^2y^3$ (or $x^3y^2$ by simply interchanging the variables). One of these cases is also studied from a different perspective in the recent work~\cite{GF1}. 

Thus, the general quintic Hamiltonian for which the Newton polygon has one interior point, is given by
\begin{equation}
\label{gen_quintic_Ham}
H^{(5)}\big(x(z),y(z);z\big)  = x^2 y^3 + a_{22}(z)\,x^2 y^2 +a_{21}(z)\,x^2 y + a_{12}(z)\,xy^2 + a_{20}(z)\, x^2 + a_{11}(z)\,xy + a_{10}(z)\,x + a_{01}(z)\, y,
\end{equation}
with analytic coefficient functions (or the same Hamiltonian with $x$ and $y$ interchanged). By a shift in $y$ we can immediately eliminate either $a_{22}$ or $a_{20}$ where here, we choose to do the latter. Thus, the general Hamiltonian from which we start our analysis is 
\begin{equation}
\label{quintic_Ham_start}
H^{(5)}\big(x(z),y(z);z\big)  = x^2 y^3 + a_{22}(z)\,x^2 y^2 +a_{21}(z)\,x^2 y + a_{12}(z)\,xy^2 + a_{11}(z)\,xy + a_{10}(z)\,x + a_{01}(z)\, y.
\end{equation}
In the following we will see that subsequent coalescences of base points in the general quintic Hamiltonian~\eqref{quintic_Ham_start} yield instances of Hamiltonians related to the (modified) Painlev\'e equations VI, V and IV: 
\begin{equation*}
\begin{array}{c c c c c}
    \includegraphics[width=.14\textwidth,valign=c]{NP_P6_quintic.pdf} & \xrightarrow[a_{21}(z)\to 0]{\text{coal.}} & 
    \includegraphics[width=.14\textwidth,valign=c]{NP_P5_quintic.pdf} & \xrightarrow[a_{22}(z) \to 0]{\text{coal.}}  & \includegraphics[width=.14\textwidth,valign=c]{NP_P4_quintic.pdf} \\[1.5ex]
    \text{\small Painlev\'e VI} & &  
    \text{\small Painlev\'e V} & &  
    \text{\small Painlev\'e IV}  
\end{array}
\end{equation*}

\subsection{A modified Painlev\'e VI equation}
There is a remaining scaling degree of freedom for $x$ and $y$ that leaves the form of the quintic term $x^2 y^3$ of the Hamiltonian~\eqref{quintic_Ham_start} invariant,
\begin{equation}
\label{quintic_scaling}
    x \to \widetilde{x} = a(z)^3 x, \qquad y \to \widetilde{y} = a(z)^{-2} y\,,
\end{equation}
which allows us to fix one of the coefficient functions. Here, we consider the consider the coefficient functions $a_{22}(z)$ and $a_{21}(z)$ as related. This is achieved by the following choice for $a(z)$,
\begin{equation}
    a(z) = \left( a_{22}(z)^2 - 4 a_{21}(z) \right)^{1/4}\,,
\end{equation}
so that the coefficient of the term $x^2 y^2$ becomes
\begin{equation}
    \widetilde{a}_{22}(z) = a_{22}(z) \cdot a(z)^{-2} = \frac{a_{22}(z)}{\sqrt{a_{22}(z)^2-4 a_{21}(z)}}\,.
\end{equation}
The coefficient of the $x^2 y$ term then becomes
\begin{equation}
    \widetilde{a}_{21}(z) = a_{21}(z) \cdot a(z)^{-4} = \frac{a_{21}(z)}{a_{22}(z)^2-4 a_{21}(z)} = \frac{\widetilde{a}_{22}(z)^2 -1}{4}\,.
\end{equation}
All other coefficient functions $a_{jk}(z)$ are also redefined in this way to coefficients $\widetilde{a}_{jk}(z)= a_{jk}(z) \cdot a(z)^{-3j+2k}$, but we drop the tildes again from here onwards. After the scaling, for convenience in the blow-up structure, we also let 
\begin{equation}\label{eq:a01_tilde}
a_{01}(z) = \frac{a_{12}(z)^2}{4}-\widetilde{a}_{01}(z)^2\,,
\end{equation}
(keeping the tilde on $\widetilde{a}_{01}$), the Hamiltonian thus taking the form
\begin{equation}
\label{quintic_Ham_P6_modified}
\begin{split} 
H^{(5)}_1\big(x(z),y(z);z\big) &= x^2 y^3 + a_{22}(z)\,x^2 y^2 + \frac{a_{22}(z)^2-1}{4} \,x^2 y + a_{12}(z)\,xy^2 + a_{11}(z)\,xy \\[1ex] 
&~~+ a_{10}(z)\,x + \left( \frac{a_{12}(z)^2}{4}-\widetilde{a}_{01}(z)^2\right) y\,.
\end{split} 
\end{equation}
We will now perform the sequence of blow-ups for the two base points 
\begin{equation}\label{eq:H5_base_points}
    p_1: (U_0,V_0)=(0,0)\,, \qquad p_2:(u_0,v_0)=(0,0)\,,
\end{equation}
of the Hamiltonian system in the extended phase space $\mathbb{CP}^2$. Each cascade of blow-ups will lead to one or more conditions on the coefficient functions in order for the equations of motion to have the Painlev\'e property, which is obtained by a cancellation of terms in the system of equations after the last blow-up of a cascade. From the first base point in~\eqref{eq:H5_base_points} the cascade splits into two branches:
\begin{equation}\label{eq:H5P6_casc1}
\setlength{\arraycolsep}{0pt}
    \begin{array}{c c c c}
        & &  & \left(U_2^{+},V_2^{+}\right)=\left(+ \widetilde{a}_{01}(z)-\dfrac{a_{12}(z)}{2}\,,0\right) \\[-1ex]
         & & ~ \shortarrow{5} &  \\[-2ex]
         p_1\colon~ (U_0,V_0)=(0,0) ~\leftarrow~ & (U_1,V_1)=(0,0) &  & \\[-.5ex]
         & & \shortarrow{3} & \\[-3ex]
         & & & \left(U_2^{-},V_2^{-}\right)=\left(- \widetilde{a}_{01}(z)-\dfrac{a_{12}(z)}{2}\,,0\right)
    \end{array}
\end{equation}
and we find the regularising conditions at the end of the last charts to be
\begin{equation}
    \frac{{\rm{d}} a_{12}}{{\rm{d}} z} \pm 2\,\frac{{\rm{d}}\widetilde{a}_{01}}{{\rm{d}} z} = 0\,.
\end{equation}
Taking the sum and difference of the expressions above, we find that both the coefficient functions $\widetilde{a}_{01}$ and $a_{12}$ are constants, and in the following we will drop their dependence on $z$.

The cascade originating from the second base point in~\eqref{eq:H5_base_points} splits into three branches with coordinates labelled by~$(u_i,v_i)$ for the first one and $(u_i^{\pm},v_i^{\pm})$ for the second and third one, respectively. 
The cascade splitting into three branches is: 
\begin{equation}\label{eq:H5P6_cascade_2}
\setlength{\arraycolsep}{0pt}
    \begin{array}{l l l}
        &  & (u_1^+,v_1^+)=\left(0,\dfrac{1-a_{22}(z)}{2}\right)~\leftarrow~ \left(u_2^{+},v_2^{+}\right)=\left(\,0\,,f^+(z)\right) \\[-1ex]
         & ~ \shortarrow{5} &    \\[-.5ex]
         p_2\colon~ (u_0,v_0)=(0,0) & ~\leftarrow~ & (u_1,v_1)=(0,0) ~\leftarrow~ \left(u_2,v_2\right)=\left(0\,,\dfrac{4\,a_{10}(z)}{1-a_{22}(z)^2} \right) \\[1ex]
         & ~ \shortarrow{3} & \\[-2ex]
         & & (u_1^-,v_1^-)=\left(0,\dfrac{-1-a_{22}(z)}{2}\right)~\leftarrow~  \left(u_2^{-},v_2^{-}\right)=\left(\,0\,,f^-(z) \right)
    \end{array}
\end{equation}
with 
\begin{equation}
    f^{\pm}(z) = \frac{\mp \! \left(a_{12}\left(\pm 1-a_{22}(z)\right)^2+4a_{10}(z)-2a_{11}(z)\,a_{22}(z)+2\,\dfrac{\d a_{22}}{\d z}\right) - 2a_{11}(z) }{2\left(\pm 1 - a_{22}(z) \right)}\,. 
\end{equation}
The three regularising conditions we find at the end of the branches are then: 
\begin{align}
    \label{eq:P6_cond1}
    &\left(1 - a_{22}(z)^2 \right)\frac{{\rm{d}}a_{10}}{{\rm{d}} z}  + 2\, a_{10}(z)\, a_{22}(z)\, \frac{{\rm{d}}a_{22}}{{\rm{d}} z} = 0\,, \\[2ex]
    \label{eq:H5_P6_cond2_cond3}
\begin{split} 
    &4 \left(\pm 1 \!-a_{22}(z)\right)\frac{\d a_{10}}{\d z} + 2\left(\pm 1 -a_{22}(z)\right)^2 \frac{\d a_{11}}{\d z} +\left(a_{12}\left(\pm 1 \!- a_{22}(z)\right)^2 -4 a_{10}(z) -2 \,\frac{\d a_{22}}{\d z}\right)\frac{\d a_{22}}{\d z}  \\[.5ex]
    &+ 2 \left(\pm 1 \! - a_{22}(z)\right) \,\frac{\d^2 a_{22}}{\d z^2} = 0\,.
\end{split}
\end{align}

The configuration of exceptional divisors for the rational surface obtained from $\mathbb{CP}^2$~\eqref{eq:CP2} blown up 11 times (at points with coordinates~\eqref{eq:H5P6_casc1}, \eqref{eq:H5P6_cascade_2}) is shown on the left of~\eqref{eq:quinticP6_surface_type}. Under two blow-downs we determine the minimal surface diagram for Painlev\'e VI:
\begin{equation}\label{eq:quinticP6_surface_type}
\includegraphics[width=.22\textwidth,valign=c]{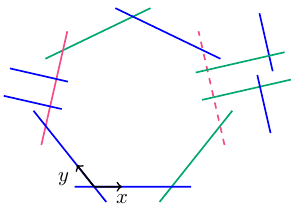}~~ \to ~~ \includegraphics[width=.2\textwidth,valign=c]{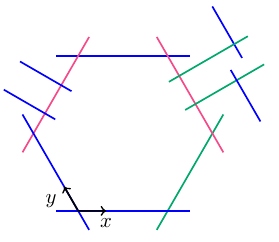}
    ~~ \to ~~ \includegraphics[width=.16\textwidth,valign=c]{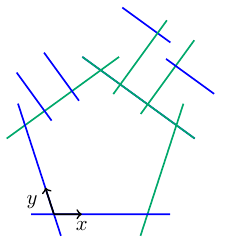} ~~\implies ~~ \includegraphics[width=.05\textwidth,valign=c]{D4.pdf}
\end{equation}
where the blue lines denote curves with self-intersection $-1$, the green $-2$, the red $-3$ and the dashed red $-4$. 
By selecting only the $-2$-curves we take the dual graph of this configuration, where lines become nodes which are connected if the corresponding lines intersect. In this way we obtain the extended Dynkin diagram for the surface type $D_{4}^{(1)}$ associated with the Painlev\'e VI equation (see page~\pageref{DDiag}).

By taking linear combinations, the two conditions~\eqref{eq:H5_P6_cond2_cond3} together with condition~\eqref{eq:P6_cond1} can be reduced to the more convenient forms 
\vspace*{-2ex}
\begin{align}\label{eq:P6_cond1_2_simpl}
    \frac{\d a_{10}}{\d z} &= \frac{2\,a_{10}(z) \, a_{22}(z) }{a_{22}(z)^2-1}\,\frac{\d a_{22}}{\d z}\,,  \qquad 
    \frac{\d a_{11}}{\d z} = \frac{a_{12}\left(a_{22}(z)^2-1\right)+4\,a_{10}(z)+\dfrac{\d a_{22}}{\d z} }{2\left(a_{22}(z)^2-1\right)}\,\frac{\d a_{22}}{\d z}\,, \\[1ex]
\label{eq:P6_cond3_simpl}
    \frac{\d^2 a_{22}}{\d z^2} &=  \frac{2\,a_{22}(z)}{a_{22}(z)^2-1}\,\left(\frac{\d a_{22}}{\d z}\right)^{\!\!2}\,.
\end{align}
The last equation~\eqref{eq:P6_cond3_simpl} admits the elementary solution 
\begin{equation}
    a_{22}(z) = \frac{1-\exp(2\,c_1(z+c_2))}{1+\exp(2\,c_1(z+c_2))}\,.
\end{equation}
We assume here that $c_1\neq 0$, as otherwise the Hamiltonian turns out to be autonomous and therefore an integral of the motion, while we are interested in the non-autonomous case. Then we can shift and re-scale $z$ such that $c_1=1/2$ and $c_2=0$, simplifying this coefficient function to
\begin{equation}
    a_{22}(z) = \frac{1-\exp(z)}{1+\exp(z)} \quad \left( = \tanh(z) \right) \,.
\end{equation}
This in turn allows us to solve the conditions~\eqref{eq:P6_cond1_2_simpl}: 
\begin{equation}
    a_{10}(z) = \frac{a \, \exp(z)}{\left(1+\exp(z)\right)^2}   \,, \qquad a_{11}(z) = \frac{a_{12}+1-a}{1+\exp(z)}+b\,, 
\end{equation}
where $a$ and $b$ are integration constants.
Thus, we have obtained a Hamiltonian system with $4$ free parameters, $a,b,\widetilde{a}_{01},a_{12} \in \mathbb{C}$,
\begin{equation}
\label{H51}
\begin{split} 
H^{(5)}_1\big(x(z),y(z);z\big) &= x^2 y^3 + \frac{1-\exp(z)}{1+\exp(z)}\,x^2 y^2 - \frac{\exp(z)}{(1+\exp(z))^2} \,x^2 y + a_{12}\,xy^2 + \left( \frac{a_{12}+1-a}{1+\exp(z)}+b \right) xy \\[1ex] 
&~~+ \frac{a \, \exp(z)}{\left(1+\exp(z)\right)^2}\,x + \left( \frac{a_{12}^{\,2}}{4}-\widetilde{a}_{01}^{\,2}\right) y\,.
\end{split} 
\end{equation} 
Eliminating $x(z)$ from the system of equations obtained from this Hamiltonian yields a second-order equation in $y(z)$. Under the further substitution
\begin{equation}
    y(z) \to \frac{y(z)}{1+\exp(z)},
\end{equation}
this becomes equation~\eqref{mP6}, which above we call the modified Painlev\'e VI equation.

We also relate this system directly to the { system derived from the} Okamoto Hamiltonian $H^{\text{Ok}}_{\text{VI}}$ in~\eqref{eq:Okamoto_H6} in the variables $(q(t),p(t))$.  
Namely, starting from the system obtained from the Hamiltonian~\eqref{H51}, by considering the following change of the independent variable $z \to z(t)$ and the dependent variables $(x(z),y(z)) \to (p(t),q(t))$:
\begin{equation}
    \exp(z) = - t\,, \qquad p(t) = \frac{x}{1-t}\,, \qquad q(t) = (1-t)\,y \,, 
\end{equation}
we derive { the Hamiltonian system associated with}~\eqref{eq:Okamoto_H6}, with the following relations for the parameters:
\begin{equation} 
\label{Okcoeff}
\begin{split}
 \widetilde{a}_{01}^{\,2}&=\frac{\kappa}{2} -\frac{(\kappa_t +\kappa_{0}+\kappa_{1}-1)^2}{8} \,, \quad a=\frac{\kappa_{0}}{\sqrt{2}} \,, \quad b=\frac{\kappa_{0}+\kappa_1}{\sqrt{2}}-1  \,, \quad a_{12} =\frac{\sqrt{3-\kappa_t^2}-\kappa_{0}-\kappa_1}{\sqrt{2}}+1 \,. 
\end{split} 
\end{equation}

\subsection{\texorpdfstring{A modified quintic Painlev\'e V system }{P5}}
Starting from the quintic Hamiltonian~\eqref{quintic_Ham_P6_modified}, we coalesce the base points labelled by $(u_1^{+},v_1^{+}) \to (u_1,v_1)=(0,0)$ in~\eqref{eq:H5P6_cascade_2}, by sending $a_{22} \to 1$, with the effect that $a_{21}(z)$ vanishes, resulting in the Hamiltonian
\begin{equation}\label{H5_P5}
    H^{(5)}_2\big(x(z),y(z);z\big)  = x^2 y^3 + x^2 y^2  + a_{12}(z)\,xy^2 + a_{11}(z)\,xy + a_{10}(z)\,x + \left(  \frac{a_{12}(z)^2}{4}-\widetilde{a}_{01}(z)^2\right)y.
\end{equation}
We consider the Hamiltonian system in the extended phase space $\mathbb{CP}^2$, with base points 
\begin{equation}\label{eq:H5_2_base_points}
p_1: (U_0,V_0)=(0,0)\,, \qquad p_2:(u_0,v_0)=(0,0) \,.    
\end{equation}
The conditions for which the Hamiltonian systems is regularised in all the charts, are as usual determined at the final step in the cascades of blow-ups. The cascade originating from the first base point in~\eqref{eq:H5_2_base_points} splits into two branches, and it coincides with the one reported in~\eqref{eq:H5P6_casc1}, hence sharing the conditions~\eqref{eq:P6_cond1}. Therefore, the coefficient functions $\widetilde{a}_{01}$ and $a_{12}$ are constant. 
The cascade emerging from the second base point in~\eqref{eq:H5_2_base_points} splits into two branches:
\begin{equation}\label{eq:H5P5_cascade_2}
\setlength{\arraycolsep}{0pt}
    \begin{array}{l l l}
        &  & (u_1^-,v_1^-)=\left(0,-1\right)~\leftarrow~ (u_2^-,v_2^-) = \big(0\,,a_{11}(z)-a_{10}(z)-a_{12}\big) \\
         & ~ \shortarrow{5} &    \\[-.5ex]
         p_2\colon~ (u_0,v_0)=(0,0) & ~\leftarrow~ & (u_1,v_1)=(0,0) ~\leftarrow~ (U_2,V_2) = (0\,,0)  ~ \leftarrow ~ \left(\widetilde{U}_3,\widetilde{V}_3\right) = (-a_{10}(z)\,,0) \\[1ex]
         &  & \hspace{50ex} \shortarrow{2} \\[-.5ex]
         & & \hspace{20ex} (U_4,V_4) = \left(\,a_{10}(z)\big(a_{11}(z)-a_{10}(z)\big)-\dfrac{\d a_{10}}{\d z}\,,0\,\right)
    \end{array}
\end{equation}
with the intermediate change of variables $\left(\widetilde{U}_3,\widetilde{V}_3\right)=(1/U_3,V_3)$, 
yielding the following regularising condition for the system in the final charts:
\begin{align}
    \label{eq:P5_cond4}
    &\frac{\d a_{11}}{\d z} = \frac{\d a_{10}}{\d z}\,,\\[1ex]
    \label{eq:H5_P5_cond3}
    &\frac{\d^2 a_{10}}{\d z^2} -\frac{1}{a_{10}(z)}\left( \frac{\d a_{10}}{\d z} \right)^{\!\!2} +  a_{10}(z)\left(\frac{\d a_{10}}{\d z} - \frac{\d a_{11}}{\d z}\right) = 0\,. 
\end{align}

The configuration of exceptional divisors of the rational surface $\mathbb{CP}^2$ blown up 11 times (at the points with coordinates~\eqref{eq:H5P6_casc1}, \eqref{eq:H5P5_cascade_2}) is shown on the left in~\eqref{eq:D5_surface_ratio}, where the blue denotes $-1$-curves, the green $-2$-curves, the red $-3$-curves and the dashed red $-4$-curves. Under two blow-downs we obtain the minimal configuration of $-2$-curves: 
\begin{equation}\label{eq:D5_surface_ratio}
    \includegraphics[width=.21\textwidth,valign=c]{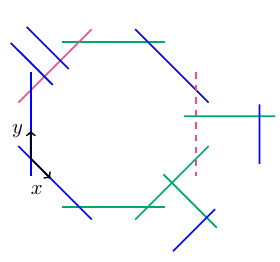}~~ \to ~~ \includegraphics[width=.20\textwidth,valign=c]{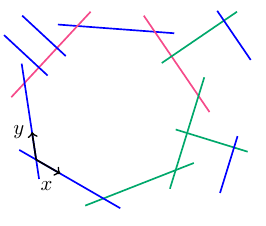}
    ~~ \to ~~ \includegraphics[width=.20\textwidth,valign=c]{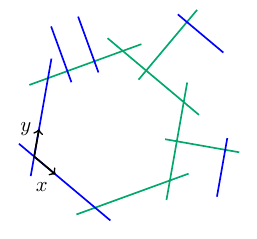} ~~\implies ~~ \includegraphics[width=.08\textwidth,valign=c]{D5.pdf}
\end{equation}
Taking the dual graph of the $-2$-curves configuration, we identify the extended Dynkin diagram $D_5^{(1)}$, associated with the Painlev\'e V equation (see page~\pageref{DDiag}).

Using  condition~\eqref{eq:P5_cond4}, the condition~\eqref{eq:H5_P5_cond3} simplifies, and can be easily solved, fixing the coefficient $a_{10}(z)$, 
\begin{equation}
    a_{10}(z)= \exp(c_1 z + c_2)\,, 
\end{equation}
with $c_1,c_2 \in \mathbb{C}$ integration constants. Considering the case $c_1 \neq 0$ we shift and re-scale $z$ so that
\begin{equation}
    a_{10}(z) = \exp(z)\,, \qquad a_{11}(z) = \exp(z)+a_{11}\,,
\end{equation}
with $a_{11} \in \mathbb{C}$ constant.
The Hamiltonian $H^{(5)}_2$ in~\eqref{H5_P5} now becomes
\begin{equation}\label{quintic_Ham_P5}
    H^{(5)}_2\big(x(z),y(z);z \big) = x^2 y^3 + x^2 y^2 + a_{12}\, x y^2 + \left( \exp(z) + a_{11} \right) x y + \exp(z)\, x + \left( \frac{a_{12}^2}{4} - \widetilde{a}_{01}^2 \right) y \,. 
\end{equation}
Under elimination of $x$ from the related Hamiltonian system, we obtain the modified Painlev\'e $\text{V}$ equation with $w(t)=y(z)-1$, $\exp(z)=t$, and the following relations for the parameters in~\eqref{mPV}:
\begin{equation}
    \alpha = 2\,\widetilde{a}_{01}^{\,2} , \quad \beta = -\frac{1}{2}(a_{11}-a_{12})^2,\quad \gamma = 1-a_{11} ,\quad \delta = -\frac{1}{2} \,.
\end{equation}
To obtain the {system derived from the} Okamoto Hamiltonian $H^{\text{Ok}}_{\text{V}}$ in~\eqref{eq:Okamoto_H5}, we implement the following changes of variables {in the Hamiltonian system derived from~\eqref{quintic_Ham_P5}} 
\begin{equation}
    \exp(z)=-\eta\, t\,, \qquad x(z)=-q(t) \,, \qquad y(z) = p(t)-1\,, 
\end{equation}
with the parameters being related via the relations
\begin{equation}
    \kappa = \frac{a_{12}^2}{4}-\widetilde{a}_{01}^2 \,, \qquad \kappa_0 = a_{12}-a_{11}\,, \qquad \kappa_t = a_{11}\,. 
\end{equation}
With these transformations, $w(t)=p(t)$ satisfies the standard Painlev\'e V equation~\eqref{Painleve5}.

\subsection{\texorpdfstring{A quintic Painlev\'e IV system}{P4}}
Under a further coalescence of base points for the Hamiltonian~\eqref{H5_P5} we obtain a Hamiltonian related to Painlev\'e IV. Namely, temporarily re-introducing the coefficient $a_{22}$ but then letting $a_{22} \to 0$, the two points $(u_1^-,v_1^-) \to (u_1,v_1)=(0,0)$ in~\eqref{eq:H5P5_cascade_2} coalesce, and the Hamiltonian becomes
\begin{equation}\label{H5_P4_simp}
    H^{(5)}_3 \big(x(z),y(z);z \big) = x^2 y^3 + a_{12}(z)\, x y^2 + a_{11}(z)\, x y + x + \left(\frac{a_{12}(z)^2}{4} - \widetilde{a}_{01}(z)^2 \right) y\,.
\end{equation}
In this case, we have used the scaling~\eqref{quintic_scaling} with $a(z) = a_{10}(z)^{1/3}$ to set the coefficient of the linear term $x$ in $H^{(5)}_3$ to $1$.
The Hamiltonian system derived by~\eqref{H5_P4_simp} evaluated in $\mathbb{CP}^2$ has the base points 
\begin{equation}\label{eq:H5P4_base_points}
p_1: (U_0,V_0)=(0,0)\,, \qquad p_2:(u_0,v_0)=(0,0)\,.     
\end{equation}
The cascade emerging from $p_1$ in~\eqref{eq:H5P4_base_points} is the same as~\eqref{eq:H5P6_casc1}, with the regularising conditions~\eqref{eq:P6_cond1}. Therefore, here as well the coefficients $\widetilde{a}_{01}$ and $a_{12}$ are constants. 

The cascade originating in $p_2$ in~\eqref{eq:H5P4_base_points} is 
\begin{equation}\label{eq:H5P6_cascade2}
   \begin{split} 
    &p_2\colon(u_0,v_0) = (0\,,0) ~ \leftarrow ~ (u_1,v_1) = (0\,,0) ~ \leftarrow ~ (U_2,V_2) = (0\,,0)  ~ \leftarrow ~ (U_3,V_3) = (0\,,0) ~ \leftarrow \\[1ex] 
    &~~~ \leftarrow ~ \big(\widetilde{U}_4,\widetilde{V}_4\big) = (-1\,,0)  ~ \leftarrow ~  (U_5,V_5) = (-a_{11}(z)\,,0)  ~ \leftarrow ~  (U_6,V_6) = \left(\frac{\d a_{11}}{\d z}-a_{12}\right),  
\end{split}  
\end{equation}
where we used the intermediate change of variables $\big( \widetilde{U}_4,\widetilde{V}_4 \big)=(1/U_4,V_4)$. In the last chart at the end of the cascade we find the
regularising condition 
\begin{equation}\label{eq:H5_P4_cond3}
    \frac{\d^2 a_{11}}{\d z^2} = 0\, ~ \implies ~ a_{11}(z) = c_1 z + c_2\,,
\end{equation}
with $c_1,c_2 \in \mathbb{C}$ integration constants. As usual, in case $c_1 \neq 0$ we shift and re-scale $z$ so that $a_{11}(z)=z$.

The configuration of exceptional devisor of the rational surface obtained from $\mathbb{CP}^2$ blown up $11$ times at points with coordinates~\eqref{eq:H5P6_casc1}, \eqref{eq:H5P6_cascade2} is shown below. 
Under $2$ blow-downs we obtain the (minimal) configuration of $-2$ curves for Painlev\'e IV: 
\begin{equation}\label{eq:E6_surf_ratio}
    \includegraphics[width=.21\textwidth,valign=c]{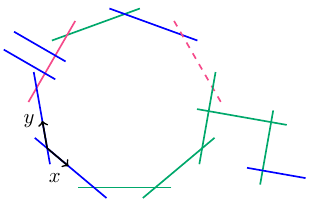}~~ \to ~~ \includegraphics[width=.205\textwidth,valign=c]{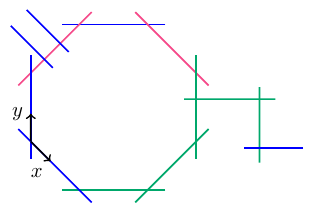}
    ~~ \to ~~ \includegraphics[width=.205\textwidth,valign=c]{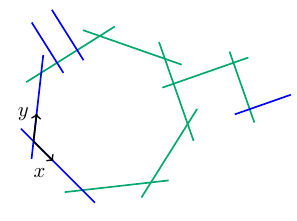} ~~\implies ~~ \includegraphics[width=.10\textwidth,valign=c]{E6.pdf}
\end{equation}
Again, the blue denotes $-1$-curves, the green $-2$-curves, the red $-3$-curves, and the dashed red $-4$-curves. 
Taking the dual graph of the configuration of $-2$ curves yields the extended Dynkin diagram of type $E_{6}^{(1)}$ (see page~\pageref{DDiag}).

Setting in~\eqref{eq:H5_P4_cond3} $c_2=0$ and $c_1=2$, the Hamiltonian~\eqref{H5_P4_simp} takes the following form:
\begin{equation}\label{quintic_Ham_P4}
    H^{(5)}_3 = x^2 y^3 + a_{12}\, x y^2 + 2z\, x y + x + \left( \frac{a_{12}^2}{4} - \widetilde{a}_{01}^{\,2} \right) y\,. 
\end{equation}
From the Hamiltonian system associated with this Hamiltonian, eliminating $x$ and $x'$ from the equations one deduces the second-order equation
\begin{equation}
    \frac{\d^2 y}{\d z^2} = \frac{3}{2y} \bigg(\!\!\left(\frac{\d y}{\d z}\right)^{\!\!2}-1\bigg) -\left(a_{12}+2 z^2-2\right)y +2\,\widetilde{a}_{01}^{\,2}\, y^3 - 4z
\end{equation}
This equation is related to the standard Painlev\'e IV in~\eqref{Painleve4} through the additional change of variables $y(z)=\dfrac{1}{w(t)}$, and the following relations between parameters: 
\begin{equation}
    \alpha = 1 - \frac{a_{12}}{2}  \,, \qquad \beta = - 2\,\widetilde{a}_{01}^{\,2} \,.
\end{equation}
{
We obtain the {system derived from the} Okamoto Hamiltonian $H^{\text{Ok}}_{\text{IV}}$ in~\eqref{eq:Okamoto_H4} by implementing the changes of variables {at the level of the system we get from}~\eqref{quintic_Ham_P4}:
\begin{equation}
     z=t\,, \qquad x(z) = -2p(t)^2 q(t)+2\kappa_{\infty}p(t) \,, \qquad y(z) = \frac{1}{p(t)} \,, 
\end{equation}
and the following relations for the parameters 
\begin{equation}
    \kappa_{\infty} = \frac{\pm 2\,\widetilde{a}_{01} - a_{12}}{4}\,, \qquad \kappa_0 = \pm 2\,\widetilde{a}_{01}\,. 
\end{equation}
Again, the variable $w(t)=p(t)$ satisfies the standard Painlev\'e IV equation~\eqref{Painleve4}.

}

\section{A septic Painlev\'e Hamiltonian system}
With our constraints for the Newton polygons for Painlev\'e equations, i.e.\ with non-zero linear terms in both $x$ and $y$ and one interior lattice point, there is one last option to investigate: the septic polynomial Hamiltonian, with general coefficient functions given by the polygon depicted for the septic case on page~\pageref{NP_H7}:
\begin{equation}\label{eq:H7}
    H^{(7)}\big(x(z),y(z);z \big) = x^3 y^4 + a_{23}(z)\, x^2 y^3 + a_{22}(z)\, x^2 y^2 + a_{12}(z)\, x y^2 + a_{11}(z)\, x y + a_{10}(z)\, x + a_{01}(z)\, y\,.
\end{equation}
For convenience in the blow-up calculations, we use the rescaling,
\begin{equation}
    x \to \widetilde{x} = b(z)^4 x, \qquad y \to \widetilde{y} = b(z)^3 y\,,
\end{equation}
with the scaling factor
\begin{equation}
    b(z) = \big( a_{22}(z)^2 - 4a_{10}(z) \big)^{1/4}
\end{equation}
to achieve that
\begin{align}
    \widetilde{a}_{22}(z) &= a_{22}(z) \cdot b(z)^{-2} = \frac{a_{22}(z)}{\sqrt{a_{22}(z)^2 - 4a_{10}(z)}}\,,\\[1ex]
    \widetilde{a}_{10}(z) &= a_{10}(z) \cdot b(z)^{-4} = \frac{a_{10}(z)}{a_{22}(z)^2 - 4a_{10}(z)} = \frac{\widetilde{a}_{22}(z)^2-1}{4}\,.
\end{align}
In $\mathbb{CP}^2$ we encounter two base points, again coinciding with the origins of the affine charts, 
\begin{equation}\label{eq:H7P6_base_points}
    p_1: (U_0, V_0)=(0,0)\,, \qquad p_2: (u_0,v_0)=(0,0)\,.
\end{equation}
The first point in~\eqref{eq:H7P6_base_points} is resolved by a cascade of blow-ups, which splits into three different branches after the second blow-up. The coordinates of the three base points after the second blow-up are given by the roots of a third degree polynomial in the coordinate $U_2$,
\begin{equation}
    U_2^3 + a_{23}(z)\,U_2^2 + a_{12}(z)\,U_2+a_{01}(z) = 0\,. 
\end{equation}
To simplify the expressions for the coordinates of these three base points, we introduce two alternative functions, $R(z)$ and $T(z)$, and redefine
\begin{align}
  a_{01}(z) &= 
  \frac{a_{12}(z)\, a_{23}(z)}{3} - \frac{2\, a_{23}(z)^3}{9} +  T(z)^{3} - R(z)^3\,, \\[1ex]
a_{12}(z) &= \frac{a_{23}(z)^2}{3}  + 3\, T(z)\, R(z)\,.
\end{align}
In these terms, the cascade emerging from the first base point in~\eqref{eq:H7P6_base_points} becomes: 
\begin{equation}\label{eq:H7P6_casc1}
\setlength{\arraycolsep}{0pt}
    \begin{array}{l l l l}
        & &  & \left(U_2^{(\overline{\omega})},V_2^{(\overline{\omega})}\right)=\left(\,\overline{\omega}\,R(z)-\omega\,T(z)-\dfrac{a_{23}(z)}{3}\,,0\right) \\[-1ex]
         & & ~ \shortarrow{5} &  \\[-.5ex]
         p_1\colon~ (U_0,V_0)=(0,0) ~\leftarrow~ & (U_1,V_1)=(0,0) &~\leftarrow~ & \left(U_2^{(1)},V_2^{(1)}\right)=\left(R(z)-T(z)-\dfrac{a_{23}(z)}{3}\,,0\right) \\[1ex]
         & &~ \shortarrow{3} & \\[-2ex]
         & & & \left(U_2^{(\omega)},V_2^{(\omega)}\right)=\left(\,\omega\,R(z)-\overline{\omega}\,T(z)-\dfrac{a_{23}(z)}{3}\,,0\right)
    \end{array}
\end{equation}
where $\omega = \frac{1}{2}(-1+i\sqrt{3})$ is a third root of unity, $\overline{\omega}$ its complex conjugate and we label the coordinates of the three base points by $1,\omega,\overline{\omega}$. The regularisation conditions coming from the blow-up calculations then become
\begin{equation}\label{condRTconst}
    \omega^j \frac{\d R}{\d z} - \overline{\omega}^j \frac{\d T}{\d z} - \frac{1}{3} \frac{\d a_{23}}{\d z} = 0,\quad (j=0,1,2) \quad \Longrightarrow \quad 
    \frac{\d a_{23}}{\d z} = \frac{\d R}{\d z} = \frac{\d T}{\d z} = 0\,,
\end{equation}
so that $a_{23}$, $R$ and $T$ are constants.

The cascade originating from the second base point in~\eqref{eq:H7P6_base_points} splits into two branches, whose coordinates are labelled with $\pm$ respectively:
\begin{equation}\label{eq:H7P6_casc2_casc3}
\setlength{\arraycolsep}{0pt}
    \begin{array}{c c c}
     & & (U_4^{+},V_4^{+})=\left(F^+(z) ,\,0\right) \\[1ex]
    &    &  \shortarrow{6} \\[.5ex]
    & & \big(\widetilde{U}_3^{+},\widetilde{V}_3^{+}\big)=\left(\dfrac{-a_{22}(z) + 1}{2}\,,\,0\right) \\[-1.5ex]
    & ~ \shortarrow{5} & \\[-1ex]
         p_2\colon(u_0,v_0)=(0,0) ~\leftarrow~ (u_1,v_1)=(0,0) ~\leftarrow~ (U_2,V_2)=(0,0) & & \\
         & ~ \shortarrow{3}  & \\[-2ex]
         & & \big(\widetilde{U}_3^{-},\widetilde{V}_3^{-}\big)=\left(\dfrac{-a_{22}(z) - 1}{2}\,,\,0\right) \\[2.5ex]
         & & \shortarrow{2} \\[.5ex]
         & & (U_4^{-},V_4^{-})=\left(F^-(z) ,\,0\right) \\
    \end{array}
\end{equation}
where we have used the intermediate change of variables $\big(\widetilde{U}_3^{\pm},\widetilde{V}_3^{\pm}\big)=(1/U_3^{\pm},V_3^{\pm})$, and with the coordinates $F^{\pm}(z)$ in the last points being 
\begin{equation}
    F^{\pm}(z) = -\dfrac{a_{23}}{2} \pm \left( \dfrac{a_{23}\,a_{22}(z)}{2} -a_{11}(z) +\dfrac{1}{a_{22}(z) \mp 1}\,\dfrac{\d a_{22}}{\d z} \right)\,.
\end{equation}
At the end of the regularisation process, we find the following differential conditions: 
\begin{align}\label{eq:H7conditions}
    &(a_{22}(z)\pm 1)^2 \left(2 \dfrac{\d a_{11}}{\d z}-a_{23}\, \dfrac{\d a_{22}}{\d z}\right) +2 \left(\dfrac{\d a_{22}}{\d z}\right)^2 -2 (a_{22}(z)\pm 1) \dfrac{\d^2 a_{22}}{\d z^2} = 0 \,. 
\end{align}

The configuration of exceptional devisor of the rational surface obtained from $\mathbb{CP}^2$ blown up $12$ times at points with coordinates~\eqref{eq:H7P6_casc1}, \eqref{eq:H7P6_casc2_casc3} is shown in~\eqref{eq:H7P6_surf_ratio} on the left. 
Under $3$ blow-downs we recognise the (minimal) configuration of $-2$ curves for Painlev\'e VI: 
\begin{equation}\label{eq:H7P6_surf_ratio}
\begin{split} 
    \includegraphics[width=.17\textwidth,valign=c]{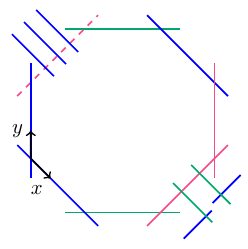}~~  &\to ~~ \includegraphics[width=.20\textwidth,valign=c]{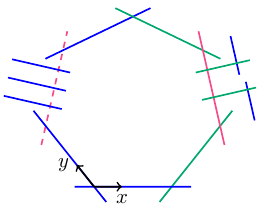}
    ~~ \to ~~ \includegraphics[width=.17\textwidth,valign=c]{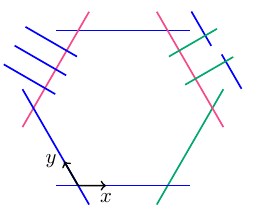} ~~   
    \to ~~ \includegraphics[width=.14\textwidth,valign=c]{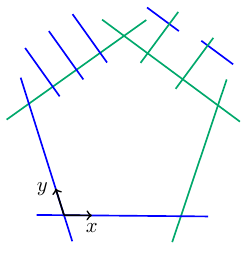} 
    ~~  \implies ~~ \includegraphics[width=.05\textwidth,valign=c]{D4.pdf}
    \end{split} 
\end{equation}
Here again the $-1$-curves are depicted in blue, $-2$-curves in green, $-3$-curves in red, and $-4$-curves in dashed red. The dual graph of the $-2$-curves configuration yields the extended Dynkin diagram $D_4^{(1)}$ (page~\pageref{DDiag}). 

Linear combinations of the two conditions~\eqref{eq:H7conditions} are recast in the more reduced forms
\begin{align}
    \label{eq:H7a11}
    \frac{\d a_{11}}{\d z} &= \frac{a_{23}}{2}\,\frac{\d a_{22}}{\d z} + \frac{1}{a_{22}(z)^2-1}\left(\frac{\d a_{22}}{\d z}\right)^{\!\!2}, \\[1ex]
    \label{eq:H7a22}
    \frac{\d^2 a_{22}}{\d z^2} &= \frac{2\,a_{22}(z)}{a_{22}(z)^2-1} \,\left(\frac{\d a_{22}}{\d z}\right)^{\!\!2}.
\end{align}
The equation in~\eqref{eq:H7a22} can be solved for $a_{22}(z)$, yielding 
\begin{equation}
    a_{22}(z) = \frac{1-\exp(2c_1(z+c_2))}{1+\exp(2c_1(z+c_2))}\,,
\end{equation}
with $c_1,c_2 \in \mathbb{C}$ integration constants. In case $c_1 \neq 0$, by shifting and rescaling the independent variable $z$, we can let $c_1=1/2$ and $c_2=0$. The coefficients functions $a_{22}(z)$ and $a_{11}(z)$ then are
\begin{equation}
    a_{22}(z) = \frac{1-\exp(z)}{1+\exp(z)} \quad \left( = \tanh(z) \right)\,, \qquad a_{11}(z) = \frac{1+a_{23}}{1+\exp(z)}+a_{11}\,,
\end{equation}
where $a_{11} \in \mathbb{C}$ is an arbitrary constant. The Hamiltonian~\eqref{eq:H7} with the explicit coefficient functions then reads as 
\begin{equation}\label{eq:H7_explicit}
\begin{split} 
    H^{(7)}_1\big(x(z),y(z);z \big) &= x^3 y^4 + a_{23}\, x^2 y^3 +   \frac{1-\exp(z)}{1+\exp(z)}\, x^2 y^2 + \left( \frac{a_{23}^2}{3}  + 3\, T R \right) x y^2 \\
    &~~ + \left( \frac{1+a_{23}}{1+\exp(z)}+a_{11} \right)  x y + x + \left(  T^{3} +a_{23}\,TR- R^3+ \frac{a_{23}^3}{9}(a_{23}-2)  \right) y\,,
\end{split} 
\end{equation}
with $a_{11}$, $a_{23}$, $R$, $T$ the four expected constant parameters.  

In the following paragraph we relate the septic Hamiltonian system to the standard Painlev\'e VI equation and match our parameters with $\alpha,\beta,\gamma,\delta$. We follow the birational coordinate transformation suggested in~\cite{KimuraMatuda1980}, adapted to our case:
\begin{equation}
    y(z) = \frac{1}{v(z)}  \,,\quad  x(z)=c\,v(z)+u(z)\,v(z)^2\,,
\end{equation}  
where $c$ is one of the following constants,
\begin{equation}
     c = \omega^j \,R - \overline{\omega}^j\, T-\frac{a_{23}}{3},\quad j \in \{0,1,2\}\,, 
\end{equation}
from which we pick the one for $j=0$ for simplicity. We obtain the {system derived from the} Okamoto Hamiltonian~\eqref{eq:Okamoto_H6} in the variables $(p(t),q(t))$ for the Painlev\'e VI equation by the following transformations {in the Hamiltonian system associated with~\eqref{eq:H7_explicit}}:
\begin{equation}
    \exp(z) = - t\,, \qquad u(z) = \frac{q(t)}{t-1}\,,\qquad v(z)=(t-1)\,p(t)\,. 
\end{equation}
With this, we find the relations connecting the parameters appearing in~\eqref{eq:H7_explicit} and the parameters $\alpha, \beta, \gamma, \delta$ in the standard form of Painlev\'e VI in~\eqref{Painleve6}, i.e. 
\begin{equation}
\begin{split} 
    \alpha &= - \frac{3}{2}(R+T)^2\,, \qquad 
    \beta = \frac{\left( 3(R-T) - a_{23} \right)^2}{18} \,, \qquad \gamma =   -\frac{1}{2}\left(  a_{11} + 1+ \frac{2a_{23}}{3} + (T-R)\right)^{\!\!2}\,, \\[1ex] 
    \delta &= -\frac{1}{2}\left(  a_{11} + 1+ \frac{2a_{23}}{3} - (T-R)\right)^{\!\!2}\,.
\end{split}
\end{equation}

\section{Conclusion}
In this paper we have obtained, for all possible degrees of a polynomial Hamiltonian, up to affine equivalence in the dependent variables, those systems for which all solutions are globally meromorphic functions. Guided by the Newton polygons of admissible Hamiltonians, we consider those polygons, with the corners $(1,0)$ and $(0,1)$ fixed, which include exactly one interior integer lattice point. Here, we find a list of $12$ different types of Hamiltonians with degrees $3$, $4$, $5$ and $7$, a priori with generic analytic coefficient functions, and, under an affine transformation, reduce them to a certain standard form. Using the blow-up procedure to obtain the space of initial conditions, we find the differential constraints on the coefficient functions such that the movable singularities of all solutions are poles. By keeping the coefficients of the highest degree terms in the Hamiltonian constant, we ensure the absence of fixed singularities in the solutions (apart from the case $D_4^{(1)}$), leading to Painlev\'e equations of type $E_j^{(1)}$, $j \in \{6,7,8\}$, or modified Painlev\'e equations of type $D_{j}^{(1)}$, $j \in \{4,5,6,7\}$. 

Where possible, we relate all the cases found to the Okamoto Hamiltonians for the Painlev\'e equations, with only a few cases that we pointed out are new in this respect. We hope that this article will serve as a reference for solving the Painlev\'e equivalence problem in the sense of Clarkson~\cite{Clarkson2019}, providing a complete list of all Painlev\'e polynomial Hamiltonian systems, up to affine transformations in the dependent variables and a possible change of independent variable of the form $t = \exp(z)$, or similar.

Finally, we summarise the results of this  work in the following, by degree of the Hamiltonians:
{\small\begin{equation*}
\setlength{\arraycolsep}{10pt}
    \begin{array}{c c c c}
    \hline \\[-2ex]
    \text{Hamiltonian deg}=7 & \text{Newton polygon} & \text{Surface type} & \text{Painlev\'e} \\[1ex]
    \hline \\[-2ex]
    \begin{aligned}
       H^{(7)}_1&= x^3 y^4 + a_{23}\, x^2 y^3 +   \frac{1-\exp(z)}{1+\exp(z)}\, x^2 y^2  \\
    &~~ + \left( \frac{a_{23}^2}{3}  + 3\, T R \right) x y^2+ \left( \frac{1+a_{23}}{1+\exp(z)}+a_{11} \right)  x y + x \\
    &~~ + \left(  T^{3} +a_{23}\,TR- R^3+ \frac{a_{23}^3}{9}(a_{23}-2)  \right) y \\[2ex]
\end{aligned}
    & \includegraphics[width=.12\textwidth,valign=c]{NP_P6_septic.pdf} &  \includegraphics[width=.05\textwidth,valign=c]{D4.pdf}   & \text{VI} \\
    \hline
    \end{array}
\end{equation*}
}
with $a_{11}, a_{23}, R,T \in \mathbb{C}$ constant parameters;
{\small\begin{equation*}
\setlength{\arraycolsep}{10pt}
    \begin{array}{c c c c}
    \hline \\[-2ex]
    \text{Hamiltonian deg}=5 & \text{Newton polygon} & \text{Surface type} & \text{Painlev\'e} \\[1ex]
    \hline \\[-2ex]
    \begin{aligned}
       H^{(5)}_1&= x^2 y^3 + \frac{1-\exp(z)}{1+\exp(z)}\,x^2 y^2 - \frac{\exp(z)}{(1+\exp(z))^2} \,x^2 y \\[1ex] 
&~~ + a_{12}\,xy^2  + \left( \frac{a_{12}+1-a}{1+\exp(z)}+b \right) xy  \\[1ex] 
&~~+ \frac{a \, \exp(z)}{\left(1+\exp(z)\right)^2}\,x + \left( \frac{a_{12}^{\,2}}{4}-\widetilde{a}_{01}^{\,2}\right) y \\[2ex]
\end{aligned}
    & \includegraphics[width=.12\textwidth,valign=c]{NP_P6_quintic.pdf} &  \includegraphics[width=.05\textwidth,valign=c]{D4.pdf}   & \text{VI} \\[1ex]
    \hline \\[-2ex]
    \begin{aligned} 
       H^{(5)}_2&= x^2 y^3 + x^2 y^2 + a_{12}\, x y^2 + \left( \exp(z) + a_{11} \right) x y \\[1ex]
       &~~+ \exp(z)\, x + \left( \frac{a_{12}^2}{4} - \widetilde{a}_{01}^2 \right) y
    \end{aligned} 
  & \includegraphics[width=.12\textwidth,valign=c]{NP_P5_quintic.pdf} &  \includegraphics[width=.07\textwidth,valign=c]{D5.pdf}   & \text{V}\\
    \hline \\[-2ex]
       H^{(5)}_3= x^2 y^3 + a_{12}\, x y^2 + 2z\, x y + x + \left( \dfrac{a_{12}^2}{4} - \widetilde{a}_{01}^{\,2} \right) y
     & \includegraphics[width=.12\textwidth,valign=c]{NP_P4_quintic.pdf} &  \includegraphics[width=.09\textwidth,valign=c]{E6.pdf}   & \text{IV} \\
    \hline 
    \end{array}
\end{equation*}}
with $\widetilde{a}_{01},a_{12},a \in \mathbb{C}$ constant parameters;

{\small\begin{equation*}
\setlength{\arraycolsep}{10pt}
    \begin{array}{c c c c}
    \hline \\[-2ex]
    \text{Hamiltonian deg}=4 & \text{Newton polygon} & \text{Surface type} & \text{Painlev\'e} \\[1ex]
    \hline \\[-2ex]
    \begin{aligned}
       H^{(4)}_1&= x^2 y^2 - \exp(2z) \left( x^2 + y^2 \right) + c\, xy \\
       &~~+ \exp(z) (a \,x + b\, y) \end{aligned}
    & \includegraphics[width=.12\textwidth,valign=c]{NP_P5_quartic.pdf} &  \includegraphics[width=.07\textwidth,valign=c]{D5.pdf}   & \text{V} \\
    \hline \\[-2ex]
       H^{(4)}_2= x^2 y^2 - x^2 + a \,xy + b\, x + \exp(z)\, y
  & \includegraphics[width=.12\textwidth,valign=c]{NP_P3_6_quartic.pdf} &  \includegraphics[width=.09\textwidth,valign=c]{D6.pdf}   & \text{III-6}\\
    \hline \\[-2ex]
       H^{(4)}_3= x^2y^2 + a \,xy + \exp(z)\left( x + y \right)
     & \includegraphics[width=.12\textwidth,valign=c]{NP_P3_7_quartic.pdf} &  \includegraphics[width=.11\textwidth,valign=c]{D7.pdf}   & \text{III-7} \\
    \hline \\[-2ex]
    \begin{aligned}
       H^{(4)}_4= x^4  + a_{21}\, x^2 y  + y^2 + a_{10}\, x + z\, y, \\ 
       a_{21} \neq \pm 2 \end{aligned} 
   & \includegraphics[width=.12\textwidth,valign=c]{NP_P2_quartic.pdf} &  \includegraphics[width=.135\textwidth,valign=c]{E7.pdf}   & \text{II}\\
    \hline \\[-2ex]
    \begin{aligned}
       H^{(4)}_5&= x^4 + 2\,x^2y + y^2 +4\,xy + z\,x \\
       &~~+ 4(2\pm\sqrt{3})\,y \end{aligned}   & \includegraphics[width=.12\textwidth,valign=c]{NP_P1_quartic.pdf} &  \includegraphics[width=.155\textwidth,valign=c]{E8.pdf}   & \text{I} \\
    \hline
    \end{array}
\end{equation*}}
with $\widetilde{a}_{10},a_{21},a,b,c \in \mathbb{C}$ constant parameters;

\vspace{2ex}

{\small
\begin{equation*}
\setlength{\arraycolsep}{10pt}
    \begin{array}{c c c c}
    \hline \\[-2ex]
    \text{Hamiltonian deg}=3 & \text{Newton polygon} & \text{Surface type} & \text{Painlev\'e} \\[1ex]
    \hline \\[-2ex]
       H^{(3)}_1= x(x-y)y + z \,xy + a_{10} \,x + a_{01} \,y   & \includegraphics[width=.12\textwidth,valign=c]{NP_P4_cubic.pdf} &  \includegraphics[width=.09\textwidth,valign=c]{E6.pdf}   & \text{IV} \\
    \hline \\[-2ex]
       H^{(3)}_2= xy^2 + x^2 + z\, x + a_{01} \,y    & \includegraphics[width=.12\textwidth,valign=c]{NP_P2_cubic.pdf} &  \includegraphics[width=.135\textwidth,valign=c]{E7.pdf}   & \text{II}\\
    \hline \\[-2ex]
       H^{(3)}_3= y^3 + x^2 + z\,y    & \includegraphics[width=.12\textwidth,valign=c]{NP_P1_cubic.pdf} &  \includegraphics[width=.155\textwidth,valign=c]{E8.pdf}   & \text{I} \\
    \hline
    \end{array}
\end{equation*}
}
with $\widetilde{a}_{10},a_{01} \in \mathbb{C}$ constant parameters.

\bibliographystyle{style}
\bibliography{thebiblio}

\end{document}